%% file: FutureNet_jpg.tex
\documentclass[12pt]{iopart}
\usepackage{iopams}
\usepackage{graphicx}
\usepackage{subcaption}
\usepackage{breqn}
\usepackage{amssymb}
\usepackage{color}
\usepackage{hyperref}

\begin{document}

\title[Global Optimization for Future Gravitational Wave Detectors Sites]{Global Optimization for Future Gravitational Wave Detectors' Sites}

\author{Yi-Ming Hu$^{1}$, P\'eter Raffai$^{2,3,4}$, L\'aszl\'o Gond\'an$^{3,4}$, Ik Siong Heng$^1$, N\'andor Kelecs\'enyi$^{3,4}$, Martin Hendry$^{1}$, Zsuzsa M\'arka$^2$, Szabolcs M\'arka$^2$}
\address{$^1$ SUPA, University of Glasgow, Glasgow, G12 8QQ, United Kingdom\\}
\address{$^2$ Columbia University, Department of Physics, New York, NY 10027, USA\\}
\address{$^3$ E\"otv\"os University, Institute of Physics, 1117 Budapest, Hungary\\}
\address{$^4$ MTA-ELTE EIRSA "Lend\"ulet" Astrophysics Research Group, 1117 Budapest, Hungary\\}
\ead{y.hu.1@research.gla.ac.uk}

\vspace{10pt}
\begin{indented}
\item[]July 2014
\end{indented}

\begin{abstract}
We consider the optimal site selection of future generations of gravitational wave detectors.
Previously, {\em Raffai et al.\/} optimized a 2-detector network with a combined figure of merit. This optimization was extended to networks with more than two detectors in a limited way by first fixing the parameters of all other component detectors. 
In this work we now present a more general optimization that allows the locations of all detectors to be simultaneously chosen.
We follow the definition of {\em Raffai et al.\/} on the metric that defines the suitability of a certain detector network.
Given the locations of the component detectors in the network, we compute a measure of the network's ability to distinguish the polarization, constrain the sky localization and reconstruct the parameters of a gravitational wave source.
We further define the `flexibility index' for a possible site location, by counting the number of multi-detector networks with a sufficiently high Figure of Merit that include that site location.
We confirm the conclusion of {\em Raffai et al.\/}, that in terms of {\em flexibility index} as defined in this work, Australia hosts the best candidate site to build a future generation gravitational wave detector.
This conclusion is valid for either a 3-detector network or a 5-detector network.
For a 3-detector network site locations in Northern Europe display a comparable flexibility index to sites in Australia.
However for a 5-detector network, Australia is found to be a clearly better candidate than any other location.
\end{abstract}

\section{Introduction}

Recent advances in technology should enable us in the near future to open a new gravitational-wave (GW) window for astronomy.
Although no signals have been detected yet, there are excellent prospects for the first detections to take place before the end of the current decade, as the `first generation' GW detectors LIGO \cite{Abbott2009} and Virgo \cite{Accadia2012} are upgraded to their `second generation' counterparts Advanced LIGO (aLIGO) \cite{Harry2010} and Advanced Virgo (AdV) \cite{Virgo2009}, with an increase of more than a factor of ten in sensitivity, which translates to an increase in detection rate of a thousand \cite{LSCrate2010}-\cite{ET2011}.
At the same time, detailed design studies for proposed future generation GW detectors such as the Einstein Telescope (ET) have recently been completed \cite{Sathyaprakash2012}, and the prospects for multi-messenger astrophysics and cosmology with such instruments have been investigated \cite{Chassande-Mottin2010} \cite{Sathyaprakash2009}.
It seems clear that the first successful detection of GW signals with aLIGO and AdV will provide tremendous impetus for the nascent field of GW astronomy, and thus generate renewed enthusiasm for the building of new and even more advanced detectors in the future.

However, the costs of building GW observatories, particularly future generation detectors, are very high \cite{Sathyaprakash2012}.
Even in the most optimistic scenarios, therefore, it seems unrealistic to expect that more than (say) half a dozen future generation GW detectors will be built in the coming decades.
Consequently, the optimal identification of sites for future GW detectors is an important issue that needs to be carefully considered.

The site selection must take into account many factors such as seismic stability and other sources of gravitational noise \cite{Coughlin2012}.
For future generation detectors, in order to achieve a further order-of-magnitude improvement in sensitivity there is a scientific motivation for constructing them underground.
This adds considerably to the cost, however, and may involve e.g. the building of extensive tunnels which may in turn place significant constraints on transport infrastructure -- all of which will contribute to the overall construction budget.
Thus choosing reliable sites for future GW detectors needs to consider a wide range of factors in addition to purely scientific constraints \cite{Raffai2013}.

However, the optimal choice of site will be significantly different when we are planning to optimizing multiple detectors instead of a single instrument \cite{Schutz2011}.
Unlike electromagnetic (EM) telescopes, which generally observe only a very small patch of sky at any time, GW detectors have an all-sky response \cite{Schutz2011} \cite{Fan2014}.
This property enables multiple GW detectors, when combined as a network, to gain improved information on the source -- including its position in the sky \cite{Cutler1994}.
In particular, while it is nearly impossible to localize the sky position of a GW source with only one detector in operation when no additional information (e.g. an EM counterpart) is available, for multiple detectors the difference in arrival time of the GW signal at each detector can be used to localize the source direction.
For a given GW event, different configurations of such a network of detectors would provide different information on its sky position \cite{Veitch2012}.
For example, multiple GW detectors that are concentrated in a small geographical area would result in very poor limits on the source's sky location, while a network of well-separated detectors would provide much tighter constraints.
These basic considerations provided a strong original motivation for the decision to build two LIGO detectors in geographically well-separated locations in the US, and more recently have informed the proposal to locate an aLIGO detector in India.

\begin{figure}[htbp]
\begin{center}
\includegraphics[width=\textwidth]{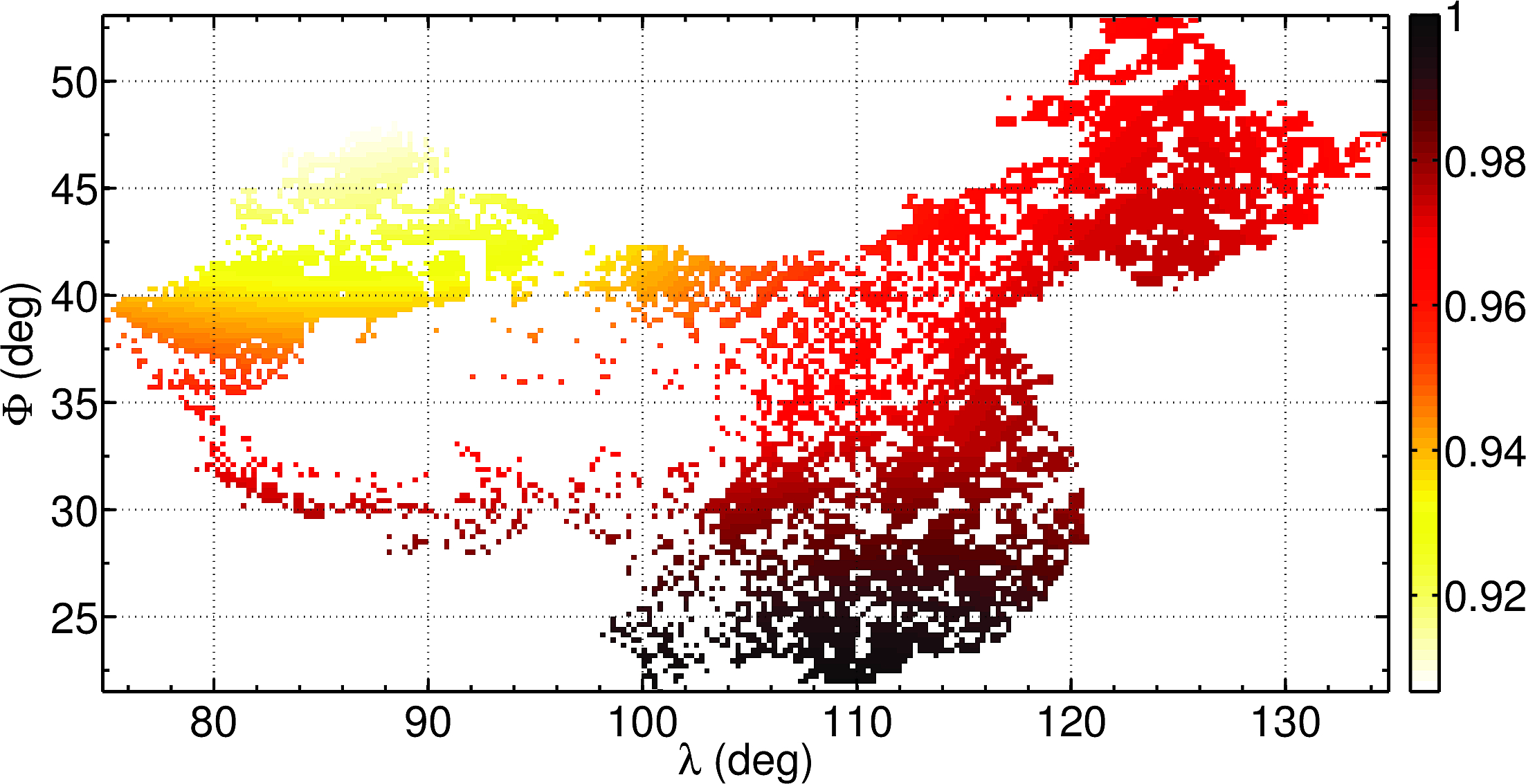}
\caption{The colormap of normalized combined metric values ($C/C_{\rm max}$), as defined in {\em Raffai et al.\/}
\cite{Raffai2013}, for various allowed geographical placements of an aLIGO-type gravitational-wave detector in China (GWD-China). The hypothetical GWD-China detector is considered as being part of a five-detector network with aLIGO Hanford, aLIGO Livingston, AdV, and KAGRA. The optimization was carried out using the same orientation angle for GWD-China for all allowed geographical locations, with one arm directing East and the other pointing North. Note that the highest values of $C/C_{\rm max}$ are found in the Southern parts of China.}
\label{fig:China}
\end{center}
\end{figure}

\begin{figure}[htbp]
\begin{center}
\includegraphics[width=\textwidth]{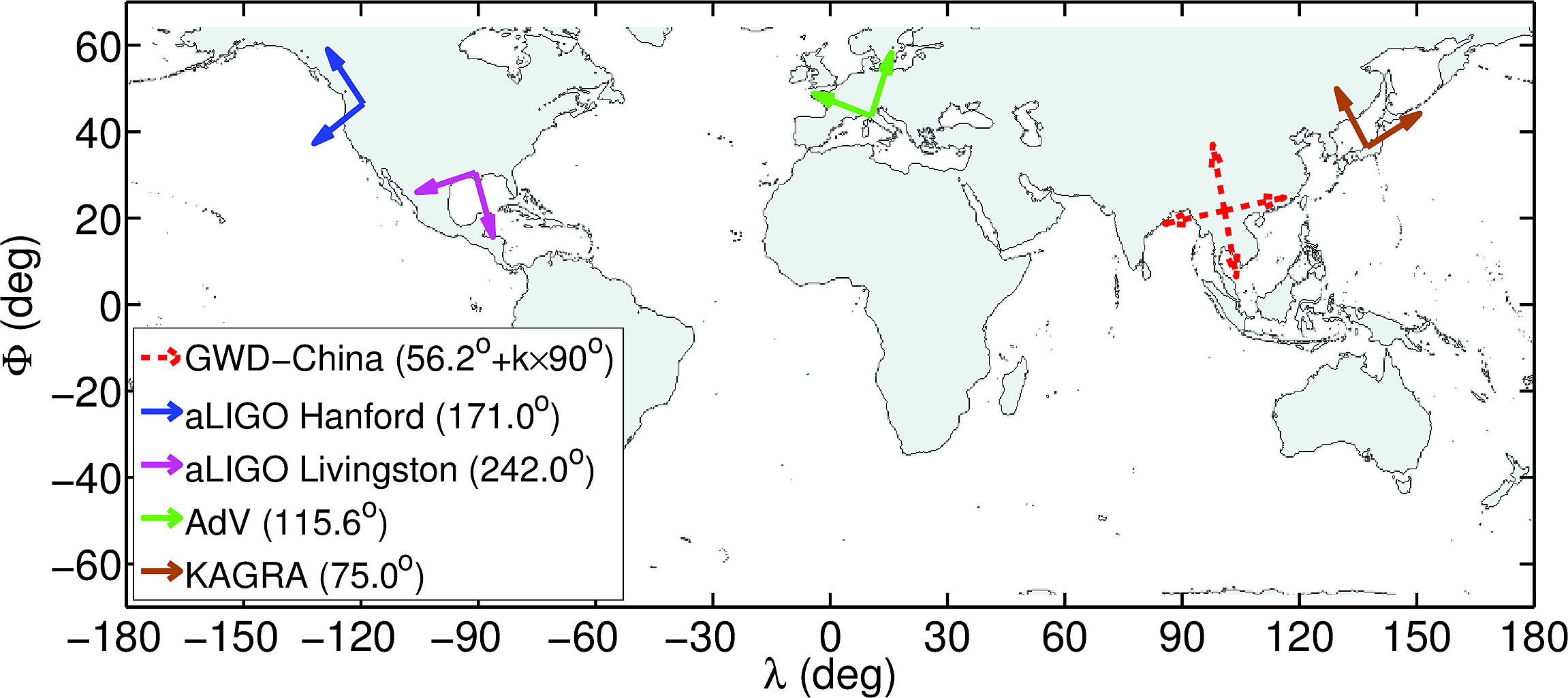}
\caption{ An example of an optimal network of advanced GW detectors including a hypothetical aLIGO-type detector in China (GWD-China). Here the optimization of the location and orientation angle of GWD-China was carried out independently; these
results would differ if they were optimized together.  In this example the Chinese detector is placed at location $[\Phi=21.75^\circ,\lambda=100.84^\circ]$, with an orientation angle (defined as the angle between the East direction and the bisector of the interferometer arms, measured counterclockwise) of $56.2^\circ +k\times 90^\circ$, $k$ being an integer. The geographical positions and orientation angles of the other detectors in the network, $([\Phi,\lambda];\alpha)$, were set to $([46.4551^\circ,-119.41^\circ];171^\circ)$ for aLIGO Hanford, $([30.56^\circ,-90.77^\circ];242^\circ)$ for aLIGO Livingston, $([43.63^\circ,10.5^\circ];115.6^\circ)$ for AdV, and $([36.42^\circ,137.3^\circ];75^\circ)$ for KAGRA. The optimization was carried out based on the normalized combined metric ($C/C_{\rm max}$) introduced in \cite{Raffai2013}. Note that the network suffers a maximum of $\sim 7\%$, $\sim 23\%$, and $\sim 2\%$ loss in terms of the $C$, $I$, and $R$ metric (as defined in Section \ref{sec:FoM}) respectively, when compared with a non-optimal orientation of GWD-China. Thus the optimization of the orientation angle is dominated by the ability of the network to reconstruct the signal polarization (as characterized by the $I$ metric).}
\label{fig:oldChn}
\end{center}
\end{figure}

\begin{figure}[htbp]
\begin{center}
\includegraphics[width=\textwidth]{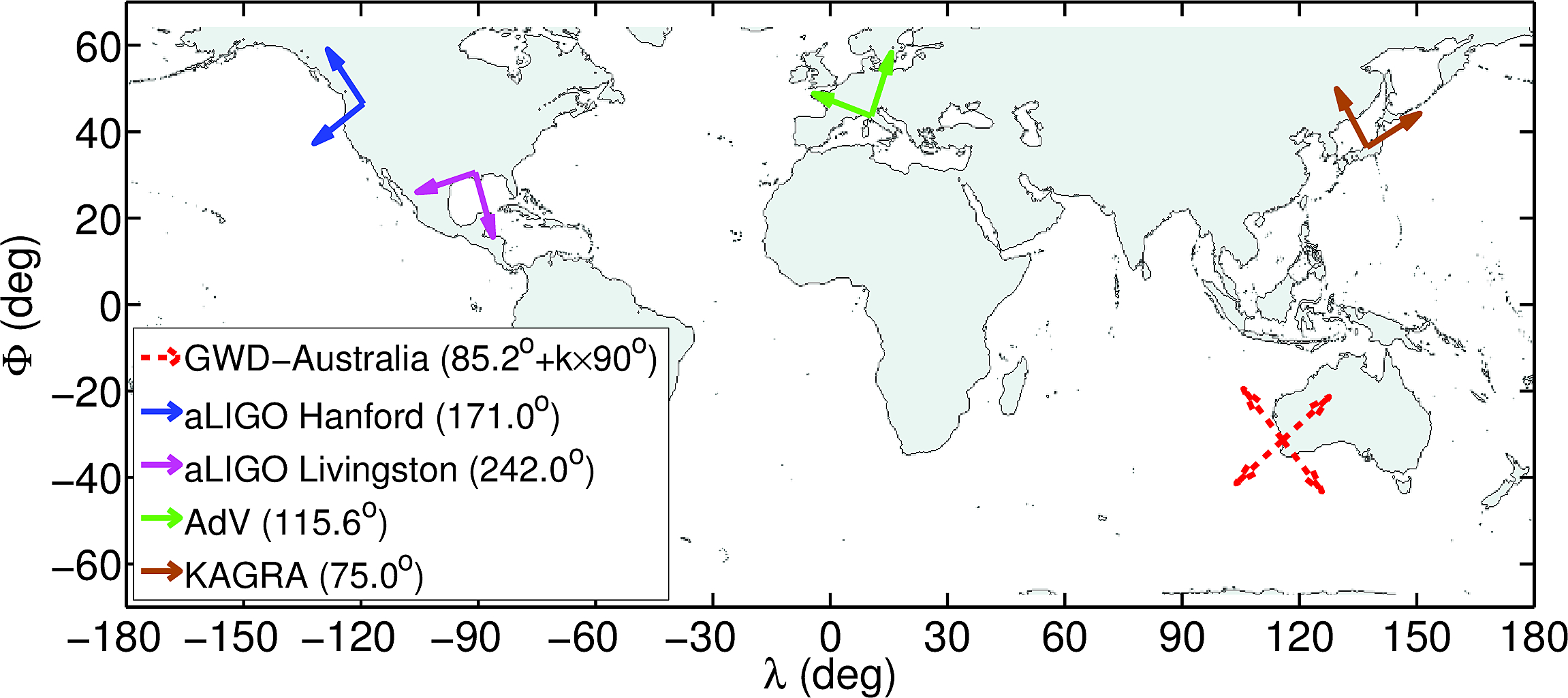}
\caption{A suggested example of an optimal network of second generation GW detectors including a hypothetical aLIGO-type detector at Gingin, Australia ($[\Phi=-31.36^\circ,\lambda=115.71^\circ]$). The optimal orientation angle for GWD-Australia (being the angle between the East direction and the bisector of the interferometer arms, measured counterclockwise) was found to be $82.2^\circ +k\times 90^\circ$, $k$ being an integer. The geographical positions and orientation angles of the other detectors in the network, $([\Phi,\lambda];\alpha)$, were set to $([46.4551^\circ,-119.41^\circ];171^\circ)$ for aLIGO Hanford, $([30.56^\circ,-90.77^\circ];242^\circ)$ for aLIGO Livingston, $([43.63^\circ,10.5^\circ];115.6^\circ)$ for AdV, and $([36.42^\circ,137.3^\circ];75^\circ)$ for KAGRA. The optimization was carried out based on the normalized combined metric ($C/C_{\rm max}$) introduced in \cite{Raffai2013}. Note that the network suffers a maximum of $\sim 8\%$, $\sim 26\%$, and $\sim 0.4\%$ loss in terms of the $C$, $I$, and $R$ metric respectively, when compared with a non-optimal orientation of GWD-Australia. Thus the optimization of the orientation angle is again dominated by the ability of the network to reconstruct the signal polarization.}
\label{fig:oldAus}
\end{center}
\end{figure}

Previously {\em Raffai et al.\/} \cite{Raffai2013} investigated constraints on optimal GW detector networks.
In that work the optimization was based on three ``Figures of Merit" -- discussed in Section \ref{sec:FoM}~-- and was mainly focused on a 2-detector network.
The generalization to networks with more detectors was carried out by adding one detector at a time, keeping the locations of all previously-sited detectors fixed.
After considering $\sim$ 1500 possible sites for the additional detector, covering all allowable regions, the best site was chosen.
This method fitted well to the aims of the {\em Raffai et al.\/} \cite{Raffai2013} study, but a simultaneous optimization of {\em all} sites in a $N>2$ detector network requires a different approach. Nevertheless, the method already proved to be very useful: for example, considering a five detector network with the first four sites to be aLIGO Livingston, aLIGO Hanford, AdV and KAGRA \cite{Somiya2012}, the optimal location and orientation of a possible fifth advanced detector in India could be determined via the exhaustive exploration of allowable five-detector configurations and the calculation of the appropriate figures of merit for the resulting network corresponding to each configuration.
This approach therefore gave indications of the relative merits of different candidate sites and orientations in the planning of aLIGO India \cite{Fairhurst2014}.
Here we have extended the method to consider the selection of a detector site in China instead of India, as shown in figures \ref{fig:China} to \ref{fig:oldAus}.

Throughout this paper, we specify the term {\em future generation GW detectors} to be ET-like detectors, {\em i.e.}, with triangular configuration, $10km$ armlength, expected to be built underground {\em etc.}, while the term {\em advanced GW detectors} refers to the detectors like the aLIGO or AdV which are being under construction or planed.
Our primary interest lies in the optimization for future generation GW detectors specifically.

In this work, we extend the method of {\em Raffai et al.\/}\cite{Raffai2013} to the case where the optimization for multiple detectors is carried out simultaneously.
This allowed us, for example, to find the optimal sites for a 3-detector or a 5-detector network, treating the locations of {\em all\/} detectors in the network as free parameters.
Also, we are interested to know if an ideal site for a detector network comprising 3 detectors could still be considered a good location when being part of a network with 5 detectors.
Moreover, we want to be able to discuss what the range of possible tolerable configurations might be.
Should the optimal configuration not be available because of any unexpected reason, such information would be priceless, and this can only be achieved by a simultaneous optimization process for the detector sites.

In practice the spatial distribution of GW sources is not isotropic, and the seismic environment varies greatly with location; however, for simplicity we ignore any such directional dependence here. 
Hence the comparison of different networks is purely determined by their relative geometric shape, and not by the physical characteristics of the sites' actual geographical locations.
Consequently shifting the whole detector network by a small amount while keeping fixed the angles between the individual detectors would not, in our analysis, affect the overall performance of the network.
Thus, we acknowledge that to ask in complete generality ``what is the optimal network?" is too ill-defined a question.
Instead, therefore, we ask a slightly different and more specific question: given the fact that we will build 3 (or 5) future generation GW detectors, where are the ideal sites for those detectors that maximise their flexibility -- i.e. such that they can belong to a large number of different ``good" detector networks?

We acknowledge, however, that at least on nearby cosmological scales the expected distribution of sources will be neither isotropic nor homogeneous -- although the impact of these effects on a given GW detector network will be mitigated by the rotation of the Earth:  a network which might be optimized e.g. to the distribution of galaxies in the Local Supercluster at one time will not in general be optimized a few hours later.
Similarly, seismic stability will also depend in detail on geographical location;  this will make some network configurations in practice more favourable than others.
However, in the following analysis, we ignore such effects for the case of simplicity.
We should clarify here that we do not take explicitly into consideration the current or planned future sites of ground based detectors.
Although these advanced detectors are expected themselves to make ground-breaking discoveries, their sensitivities are still expected to be an order of magnitude or more lower than that of proposed future generation GW detectors.
Thus, in the context of the future generation GW detector network that we are investigating in this study, such advanced detectors would make only a limited contribution.
Also, most of the current advanced detector sites are not considered ideal according to our exclusion criteria, as discussed in section \ref{sec:exc}.
As the future generation detectors are expected to be constructed underground, there is no compelling motivation to locate them at the sites of the current detectors.
Hence in this study we simply ignore the locations of current or planned advanced detectors.

\section{Figures of Merit for GW Detector Networks}\label{sec:FoM}
%Introduce the figures of merit that are used in this project.
%Refer to the CQG paper.

In {\em Raffai et al\/}\cite{Raffai2013} the authors comprehensively investigated different GW detector network configurations, characterizing their relative performance using some reasonable Figures of Merit (FoMs). So that we can conveniently compare our results with that previous work, we mostly adopt the same definitions as \cite{Raffai2013} and consistently construct our FoMs as
\begin{itemize}
\item $I$, which measures the network's capability of reconstructing the source polarization,
\begin{equation}\label{eq:I}
I = \Big(\frac{1}{4\pi}\oint \!\big|F^{{\rm network}}_+(\Phi,\lambda)-F^{{\rm network}}_\times(\Phi,\lambda)\big|^2{\rm d}\Omega\Big)^{-\frac{1}{2}}
\end{equation}
where $F^{{\rm network}}$ is the antenna factor of a network as defined in \cite{Raffai2013}.
\item $D$, which measures the accuracy with which the network can localize the sky position of the source.
\begin{equation}\label{eq:D}
D = \frac{1}{4\pi}\oint \!H\big(S-A_{90}(\Phi,\lambda)\big){\rm d}\Omega ,
\end{equation}
where $H(x)$ is the Heaviside function, $S$ is the preset threshold that we will discuss later, and $A_{90}(\Phi,\lambda)$ is the $90\%$ confidence localization region for a source located at sky position $(\Phi,\lambda)$.
\item $R$, which measures the accuracy with which the network can reconstruct the parameters of a standard compact binary source.
\begin{equation}\label{eq:R}
R = \Big(\frac{1}{4\pi}\oint \!\sigma_{\mathcal M}(\Phi,\lambda)^2{\rm d}\Omega \Big)^{-\frac{1}{2}},
\end{equation}
\end{itemize}

In the case of the $D$ metric introduced in \cite{Raffai2013}, we have used a slightly different expression for characterizing the accuracy of a GW detector network in source localization.
Our new $D$ metric is based on the method introduced in Fairhurst 2010 \cite{Fairhurst2010}, since it gives the localization error, which is a more interesting and has a more direct physical meaning than the previous definition of $D$.
Here the localization accuracy for sources at various sky positions is expressed as the angular areas of the $90\%$ confidence localization regions (ellipses) obtained by triangulation of the source after successful individual detections with $N$ GW detectors ($N>2$).
For a given $N$-detector configuration, we first calculate the localization ellipses' angular areas, for every given direction, then the $D$ metric value is computed as the percentage of the sky for which the area of these ellipses falls below a specified threshold, $S$. 
Using the simplification that all detectors in the network register the incoming GW signal with the same timing accuracy, we can directly express $S$ in ${\rm deg}^2$.
The actual choice of thresholds for $S$ that we adopted will be explained in detail in \ref{sec:opt}. 
Notice that the calculation of $D$ should include the timing uncertainty and, according to \cite{Fairhurst2010}, that value is related to SNR.
Here we assume all signals have a timing accuracy that corresponds to an SNR of 8.
This influence of this value on $D$ is degenerate with $S$ and the actual choice of value could be somewhat arbitrary.

In practise we find out that the new definition of D is more realistic.
However, it is worth mentioning that although the definition of $D$ has been changed, when considering the optimal configurations, both new and old definitions of $D$ give very similar results, suggesting an intrinsic consistency between both definitions.  So in particular for configurations with high $D$, the new definition makes negligible change to the conclusion.

Note also that in what follows we apply equal weights to each of the three individual FoMs.
This approach is a gross simplification, and ignores the possibility that the different FoMs may have different scientific purposes and their actual relative importance would depend on the context in which the detector network was operating.
For example, in many proposed applications of GW astronomy a key consideration is to identify an EM counterpart of the GW source in order, e.g., to determine its redshift or some other astrophysical characteristic of the source that is crucial for its exploitation.
In this scenario it might be that optimizing the sky localization is the most important of the three FoMs, while in other circumstances it might be that optimizing the measurement of the source polarization (which is important, for example, to break degeneracies in the sky position, or to help detecting unmodelled Burst waveforms) would be prefered.
One could straightforwardly adapt our method to such a case simply by adjusting the relative weighting of the FoMs.
Furthermore, there may be some intrinsic correlation between the FoMs, which would again call for a more generalized combined statistic that directly takes this into account.
We defer such extensions until future work, however, and address here only the case of equally weighted, fully independent, FoMs.
However, we recognize that an inappropriate choice of weights could lead to a selection criterion that was not physically motivated.

For each FoM $I$, $D$ and $R$, we find the maximum values $I_{\rm max}$, $D_{\rm max}$ and $R_{\rm max}$ and normalize the FoM to these maxima.
The squared sum of these normalized values will then give the total FoM, $C$, as defined in \cite{Raffai2013}
\begin{equation}
C = \sqrt{\Big( \frac{I}{I_{max}}\Big)^2+\Big(\frac{D}{D_{max}}\Big)^2+\Big(\frac{R}{R_{max}}\Big)^2}.
\end{equation}
With such a definition, the total FoM is not biased towards any metric.

\section{Description of Method}
As briefly discussed in the previous sections, in {\em Raffai et al.\/}\cite{Raffai2013} the network optimization is achieved by first fixing the other detectors' locations and then finding (using the FoM) the optimal site for an additional detector.
Moreover the number of detectors considered in \cite{Raffai2013} was limited to 3 in a few example cases of a global optimization.
Readers are reminded that the primary motivation for the current work was to extend the earlier results to the case where all detectors could be optimized simultaneously.
This would, of course, as a consistency check allow us to compare our results with those of \cite{Raffai2013}, as well as possibly to identify new, optimal sites.

One cannot reliably predict the actual number of GW detectors that will operate in the future since there are so many uncertainties. 
Once again, we note that we restrict our considerations to only those of future generation detectors.
Optimistically, if an unexpected and exciting new discovery were to occur with the nascent advanced detector network, then -- combined with a healthy global economic environment -- it seems reasonable that this would boost the case for building several more detectors and pushing their design envelope to the future generation.
On the other hand if the actual rate of astronomical events observed by the advanced network is at the pessimistic end of current predictions, then -- if it were combined with difficult economic circumstances -- this might strongly limit the number of future generation GW detectors that would actually be built.
Ideally, therefore, we would want the optimally-chosen sites for proposed future detectors to be as flexible as possible, so that in particular their scientific performance could be high in both optimistic and pessimistic scenarios.
Consequently, in this work we also extend our analysis to consider a 5-detector-network, in order to answer the specific question of whether a ``good'' site for a 3-detector-network is still attractive when a 5-detector-network is considered.
Thus we want to determine the optimal location for the first site so that it leaves the future the most flexibility.
Of course it is still natural to expect that the optimal sites for a 3-detector-network, supplemented by two additional, optimally-chosen sites, would be at least slightly different from the optimal sites determined simultaneously for a 5-detector-network.  However, in the event that not all new detectors would be funded and built at the same time, it may be very likely that a 3- or 4-detector network would be built first and then extended by one or more additional detectors (exactly analogous to the current proposal for  LIGO India).
Our question about assessing the performance of different detector sites as the size of the network changes would, therefore, seem to be both timely and appropriate.

\subsection{Methodology}

It is clear that, in order to optimize all detectors simultaneously while exploring the situation for a network of up to 5 detectors, the method adopted by \cite{Raffai2013} would not be appropriate.
In \cite{Raffai2013}, the geographical regions that are suitable as detector sites were reduced to $\sim1,500$ discrete candidate locations and the optimization was achieved by an exhaustive search over all of these candidates. 
The computing overhead for this approach is tolerable if one optimizes for only one site at a time.
However, if we allow even 3 detectors' locations all to be free parameters over which to be searched, then we have $\sim1,500^3\approx3.4\times10^9$ different combinations to explore.
Extending to a 5-detector-network would increase this to $\sim 7.5\times 10^{15}$ combinations -- which is far beyond what is currently realistic.
However, equally clearly, there will be a significant fraction of these combinations that correspond to networks with a relatively low FoM in which we are not really interested.
The question then becomes: how can we efficiently explore only those regions with high FoM, even for networks with a large number of detectors?
Bayesian inference methods like Markov Chain Monte Carlo (MCMC) \cite{Metropolis1953}\cite{Hasting1969}\cite{Gregory2005} or Nested Sampling \cite{Skilling2004}\cite{Skilling2006}\cite{Sivia2006} are designed to deal with such problems, and they perform especially well when the parameter space has a high dimensionality.
We therefore adopt a Bayesian inference approach here, and assign as the posterior some monotonic function of the total FoM.

For an equilateral triangular shaped interferometr, as is being proposed for the future generation GW detector \cite{Punturo2010}, the local antenna response for a given signal is nearly independent of azimuth.
Consequently we do not consider the orientation of the detectors in our network, only their geographical location, i.e. longitude and latitude.
Thus, if the detector network consists of $N$ detectors, then the dimensionality of the problem will be $2N$.
For even a 3-detector network, then, which corresponds to a 6-dimensional parameter space, we can expect that our Bayesian method will significantly outperform the brute force grid-based search.

However, one should realize that the problem we are studying is strictly {\em not\/} a Bayesian parameter estimation problem.
We are merely taking advantage of some of the technology that has been developed to carry out efficient sampling of high-dimensional Bayesian posterior distributions, and adapting our problem so that this technology may be directly applied to it.
Essentially we are only interested in efficiently identifying and sampling from regions with high FoMs.
Consequently constraints like {\em detailed balance\/}, which in general are required in MCMC applications in order to ensure that the samples are indeed drawn from the appropriate posterior distribution \cite{Gregory2005}, will be of less concern to us here.

There is a vast literature on MCMC methods, including a growing list of example applications in the field of GW astronomy \cite{vanderSluys2008}\cite{Raymond2009}\cite{Farr2013}.
The interested reader is referred to \cite{Gregory2005} \cite{Sivia2006} for a thorough overview and introduction to MCMC methods; here we present very briefly the essential principles as a precursor to introducing (in the next sub-section) a new adaptation of MCMC that is specifically tailored to our network optimization problem.

The simplest implementation of MCMC is the so-called Metropolis algorithm \cite{Metropolis1953} which we present here for the case of a posterior distribution characterized by a set of parameters $\boldsymbol{\theta}$.  To implement the Metropolis algorithm requires the following basic steps:
\begin{enumerate}%[i]
  \item Sample a random point $\boldsymbol{\theta}_{i}$ in the parameter space and evaluate the corresponding posterior, $f(\boldsymbol{\theta}_{i})$. 
  \item Propose a candidate point $\boldsymbol{\theta}^*$ distinct from the previous point $\boldsymbol{\theta}_{i}$, by sampling from a specified {\em proposal density\/}, and evaluate the posterior $f(\boldsymbol{\theta}^*)$ at the candidate point.\label{2}
  \item Calculate the Metropolis ratio $r=\frac{f(\boldsymbol{\theta}^*)}{f(\boldsymbol{\theta}_{i})}$.  Accept, with probability $\min{[r,1]}$, $\boldsymbol{\theta}^*$ as the next point, $\boldsymbol{\theta}_{i+1}$, in the chain.  If  $\boldsymbol{\theta}^*$ is {\em not\/} accepted then set 
$\boldsymbol{\theta}_{i+1}=\boldsymbol{\theta}_{i}$ .\label{3}
\end{enumerate}
Steps (\ref{2}) and (\ref{3}) are then repeated until the sampled distribution of points is regarded as having converged, and thus can be taken as representative of the actual posterior distribution.
In particular step (\ref{3}) is crucial to maintain what is referred to as {\em detailed balance\/}; it is this step that ensures that the generated chain is a sample from the desired posterior distribution.
The choice of proposal density is completely arbitrary as long as it guarantees ergodicity, although a judicious choice may greatly improve the efficiency and speed with which the algorithm converges to the desired posterior.

\subsection{Introduction to mixed MCMC}
As noted earlier, one interesting feature of our approach is that we can shift the entire detector network while keeping fixed the angles between the individual detectors without altering the FoMs that are computed, provided all sites remain within allowable regions.
As also noted earlier, including extra information about the actual anisotropic distribution of cosmological sources would break this degeneracy, but such an extension is beyond the scope of the current study.

However, another feature of our optimization problem is that it is naturally multi-modal, i.e., it contains multiple local maxima.
The territory of the Earth is divided into isolated continents and this naturally leads to significant discontinuities in the computed FoMs, thus making the distribution of optimal networks intrinsically multi-modal.
The simplest MCMC methods generally become clumsy and less reliable when dealing with posterior distributions that have multiple modes.
Hence we have developed an MCMC-based approach that still performs well for multi-modal distributions.
More specifically we have taken advantage of MCMC methods' ability to concentrate sampling in regions of high FoM while simultaneously being able to `swap' between multiple, distinct modes.
To meet these requirements we have developed a new variant of MCMC, known as {\em mixed MCMC}, that can sample independently from different regions simultaneously.
Other Bayesian sampling methods like parallel tempering MCMC \cite{Swendsen1986}, affine invariant MCMC \cite{Farr2013} and MultiNest \cite{Feroz2009} \cite{Feroz2013} are able to sample from multiple regions, but mixed MCMC is among the most efficient of such methods.

We describe fully our mixed MCMC method in {\em Hu et al. \/} \cite{Hu2014}, but here sketch out the general principles.
The basic idea is that when multiple modes are known to exist in the parameter space, the properties of `normal' Metropolis MCMC should enable the sampler to explore neighbouring regions efficiently, while maintaining a `global' communication between the samples generated in these different regions so that the sampling results can reflect the respective weight of the different modes. 
We know of the existence of multiple modes, and we assume that the location of the local maxima are also known, so that we can compute the difference vector $\boldsymbol{r}_{mn}$ connecting mode $m$ and mode $n$.  The MCMC algorithm remains the same as for a single mode in the case where the proposed candidate is generated in the same mode as where the previous point was located. However, a swap between different modes might be proposed, say, from mode $m$ to mode $n$. In that case the candidate will be shifted by $\boldsymbol{r}_{mn}$, in addition to its position being sampled from the appropriate proposal density (e.g. a multivariate Gaussian random vector) within mode $n$.

As mentioned before, the Metropolis ratio is the criterion used to determine the acceptance or rejection of a proposed candidate point.  The formalism for this is given in equation \ref{eq:correct}.
Suppose that in the $i^{th}$ step, the candidate $\boldsymbol{\theta}^*$ is proposed in the mode labeled as $n$, while the previous step locates the $i-1^{th}$ point in the mode labeled as $m$.
For each mode, a so-called {\em picking up probability\/} $p$ is assigned.
So $p^{(m)}$ represents the probability to propose a candidate that remains in the same mode, while $p^{(n)}$ is the probability that the candidate will be in the mode labeled as $n$, and the picking up probability should normalize such that 
$\sum p^{(i)} =1$.
\begin{equation}\label{eq:correct}
        r=\frac{f(\boldsymbol{\theta}^*)}{f(\boldsymbol{\theta}_{i-1})}\frac{p^{m}}{p^{n}} 
\end{equation}
We can see, therefore, that mixed MCMC will return to the normal Metropolis MCMC when the candidate point and the previous point are located in the same mode, as we expected.

The value of picking up probability should strictly follow the partition of \emph{evidence} in each mode in order for the algorithm to be optimized from a Bayesian point of view.
However, here we are making use of the novel machinery of mixed MCMC solely in order that we may sample efficiently from different distinct regions -- i.e. different modes in our parameter space.
Thus we can afford to tolerate a formal deviation from detailed balance -- and hence from exact partitioning by the evidence -- in return for increased computational efficiency and convenience. We return to this point below.

\subsection{The actual realization}\label{sec:realis}

The definition of the FoM for a network of detectors given in the previous section is not sufficiently discriminatory from the point of view of implementing MCMC.
In particular, from the way in which the equations are set up the FoM for the least optimal network differs by only a factor of two from the most optimal network.
If we define the effective `posterior', namely $f(\boldsymbol{\theta})$ in equation \ref{eq:correct}, to be simply proportional to the FoM, therefore, the MCMC sampler will waste a great deal of time in uninteresting regions of the parameter space.
Consequently, we manually set the effective `posterior' to be the exponential of the FoM, 
so that there is much greater differentiation between the least and most optimal networks.
This in turn ensures that the sampler will spend more time exploring regions with high FoM.  

In order to begin sampling, we manually partition the parameter space in a conservative manner so that one region is allowed to host at most one major mode.
The details of how this is done are discussed in \ref{sec:part}.
However, we note that the parameter space is not continuous within each region even after partition.
In order to avoid any adverse impact of discontinuity, we enable the network to contain sites that are located in unfavored regions like oceans so that the sampler can traverse between discontinuous regions.
The exclusion of disfavoured regions is discussed in more detail in section \ref{sec:exc}.
For every such `bad location' site, the posterior will be divided by the base of natural logarithms, $e$, so that the sampler does not waste too much time exploring undesired regions.
Under this formulation disfavoured regions can therefore be sampled, but are not favoured, and the discontinuity problem disappears.
Notice, moreover, that when we present our conclusions about optimal sites, those configurations that contain bad locations will in any case be automatically discarded.

Multiple CPUs were used to sample several network configurations simultaneously, in order to further boost the efficiency.
All of the samples generated were combined into 4 groups, and the sample results from these 4 groups were constantly monitored.
Once the convergence criteria were met, i.e. the properties of the 4 groups were sufficiently similar, the sampling process was stopped.

We apply an automatic stopping criterion for the convergence of mixed MCMC chain.
In each sub-chain, this convergence criterion is checked using the well-established Gelman-Rubin criterion\cite{Gelman1992} on every parameter.
Interested readers are referred to \ref{sec:opt} for details about the Gelman-Rubin criterion.

\subsection{Exclusion of unsuitable regions}\label{sec:exc}
In order to filter out regions where it would be unsuitable to build a detector, criteria similar to those in \cite{Raffai2013} were used to exclude such sites.
First, to build a detector underwater would be far from realistic due to the huge expense to build and maintain it, so we exclude all oceans, seas and continental lakes.
Furthermore, we exclude all coastlines that are within $\sim100$ km distance from the ocean, so that the micro-seismic noise due to oceanic waves is mitigated.
Similar considerations of transport and the convenience of maintenance lead to the exclusion of polar regions, regions with slope steeper than $5^\circ$, as well as regions with an elevation higher than 2000 meters above sea level \cite{agwed}.
Also, routine human activity in centres of population would induce a gravity gradient noise, so the detectors should be built far away from densely populated regions, including major roads in North America.
This exclusion is achieved by excluding regions where there is significant artificial illumination during the night, in addition to populated areas as defined according to the Natural Earth database \cite{natear} \cite{night}.
In addition, and differently from \cite{Raffai2013}, we further exclude protected areas, since undertaking construction in such regions would be generally illegal \cite{prot}.

\begin{figure}[htbp]
\begin{minipage}[c][15cm][t]{\textwidth}
%  \vspace*{\fill}
  \centering
  \includegraphics[width=\textwidth]{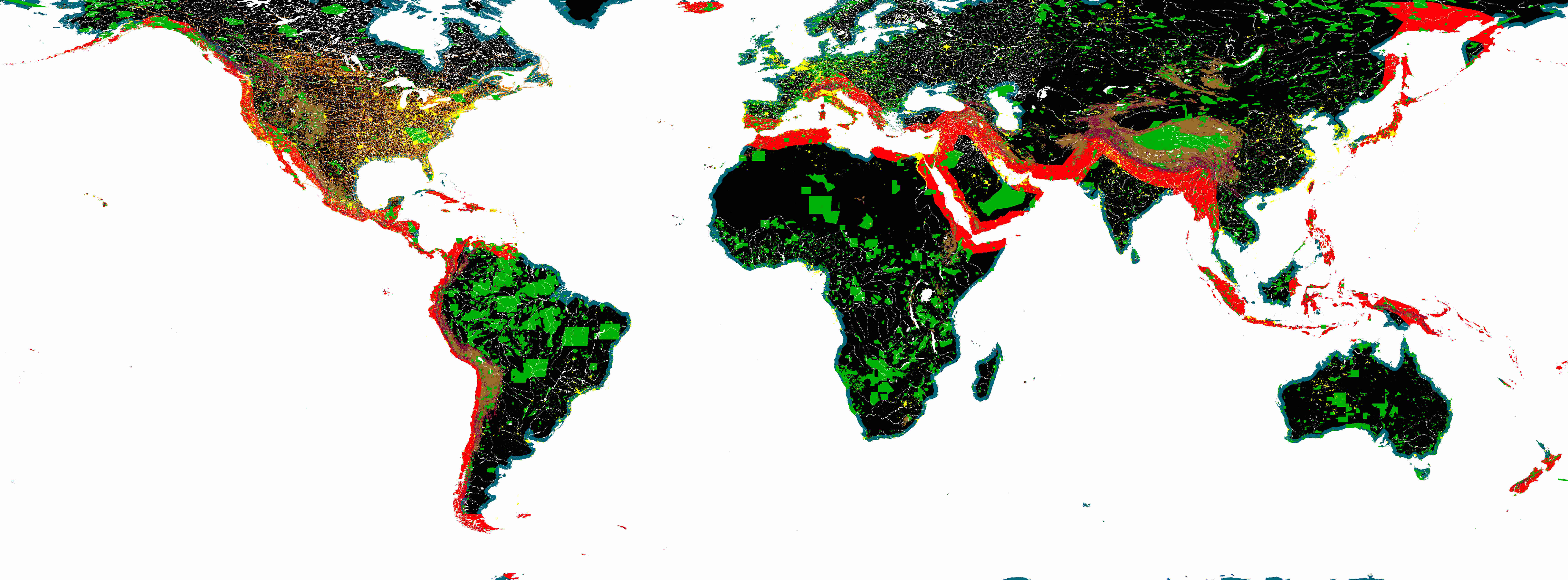}
  \subcaption{ }
  \label{fig:exclude_color}\par\vfill
  \includegraphics[width=\textwidth]{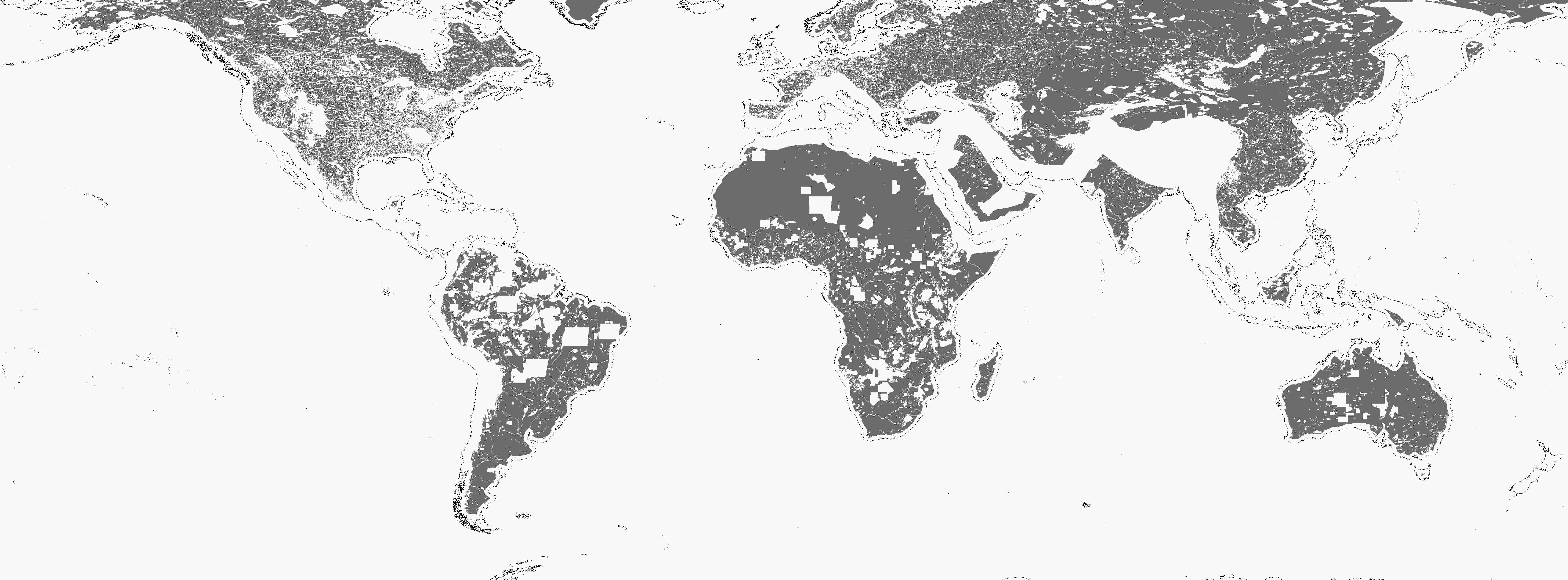}
  \subcaption{ }
  \label{fig:exclude_pale}
\end{minipage}
\caption[caption]{The figures show the exclusion of unsuitable regions according to various exclusion criteria, including a colored version and a monochromatic version.
Interested readers are encouraged to check out the full resolution files at: \url{http://elysium.elte.hu/~praffai/geomap.html}
Detailed conditions and corresponding colors (for the colored version) are listed below. 

\ref{fig:exclude_color}:
Dark green: coastlines with $\sim~100$ km width.
Red: seismically active areas with a 200 km width \cite{tectonic}.
White: oceans, seas and fresh waters, according to the Natural Earth database \cite{natear}.
Dark brown: roads of North America \cite{natear}.
Brown: elevated areas, \emph{i.e.} areas above 2000 m, according to ASTER GDEM Worldwide Elevation Data map \cite{agwed}.
Claret: high gradient areas, \emph{i.e.} areas with higher than $5^\circ$ slope \cite{agwed}.
Green: protected areas like national parks, according to the World Database on Protected Areas \cite{prot}.
Yellow: populated areas, including bright areas on the NASA night lights map \cite{night} and other populated areas like those in India and China \cite{natear}.
Black: potentially suitable regions.

\ref{fig:exclude_pale}:
White: outfiltered regions.
Grey: potentially suitable regions.
The coastlines are marked with a solid black contour.
}
\label{fig:exclude}
\end{figure}

This work used the exclusions as mentioned above.
For future work, more constraints have been proposed, such as seismically unstable regions which are not suitable because of their high level of environmental noise \cite{tectonic}.
We show our most up-to-date exclusion figure in \ref{fig:exclude}. 
Beyond that, more factors can be considered, like military regions, which might not be feasible to access; regions with active mining activity, or which contain rich mines, that could induce gravity gradient noise to the detector data.
Contaminated areas and nuclear test sites are also not ideal places.

\section{Results}\label{sec:result}
The principal goal of this work is to determine the optimal sites on which to build a network of future generation GW detectors.
However, as noted earlier, the formulation of our FoM is invariant under translations of the entire network across the Earth's surface, provided that the network shape remains unchanged and all sites remain located in allowable regions.

In order to better distinguish good sites and good network configurations from others, we can define the ``{\em flexibility index}" of a site -- \emph{i.e.}, if that site is included in our network how many possible distinct network configurations could there be that would each give a high FoM?
In this sense, a ``better'' site means one that would give more freedom, or flexibility, in choosing the locations of other detector(s).
Our results below, therefore, illustrate on a world map the flexibility index of different sites for networks containing different numbers of detectors.

We run our optimization code separately for 3-detector and 5-detector networks.
The mixed MCMC method is used so that the sampler can be concentrated in separate regions of high FoM.
Once the sampling is terminated, when the convergence criteria are met, we assume that all regions of interest should by then have accumulated sufficient samples to represent adequately the underlying distribution.
Although we want the maps to reflect the globally optimal network configurations, for networks containing different numbers of detectors, we should bear in mind the other external factors that will determine in reality where future generations of detectors are constructed.
Thus we adopt a threshold on the FoM, $90\%$ of the highest FoM, and consider {\em all\/} networks that exceed this threshold.
We then determine our flexibility index for each site by counting the number of different networks including that site which exceed our chosen FoM threshold.

For displaying our results we adopt a standard world map with $1520\times759$ pixels.
For each pixel, we convert the pixel location into geographical coordinates, and identify all the networks that contain one site within a certain distance ($\sim 200$ km for the zoom-out figures like Figure \ref{fig:CD3} and \ref{fig:CD5}, and $\sim64$ km for the zoom-in figures like Figure \ref{fig:zoom3_90}, \ref{fig:zoom3_80} and \ref{fig:zoom5_80}) of that pixel. 
This choice leads to a larger fluctuation for the zoom-in maps.

To avoid multiple counting of `similar' configurations, we determine the inner product of the unit vectors constructed from the coordinates of the sites that comprise two networks that we wish to compare.
For a network that consists of $t$ detectors, we set $\mathbf{N} = (\Phi_k,\lambda_k,\ldots,\Phi_l,\lambda_l)$, while $k,\ldots,l$ is permuted over $1,\ldots,t$.
The normalized vector $\mathbf{n} = \frac{\mathbf{N}}{|\mathbf{N}|}$, and we define the inner product of two networks $\mathbf{n}$ and $\mathbf{m}$ to be $s={\rm max}(\mathbf{n},\mathbf{m})$ that has been maximized over permutation.
We define the two networks as duplicates if their inner product is larger than 0.95; in this case one of the network configurations is discarded.
The adoption of this approach ensures that our results are not adversely affected by the somewhat arbitrary partitioning of the world map, while at the same time allowing our mixed MCMC approach to benefit from the fast identification of regions of interest precisely through use of this partitioning.
Since this inner product criterion of 0.95 is applied to unit vectors, it is independent of the number of detectors in our network.
This is very useful as it allows us to compare directly and straightforwardly our results for 3-detector and 5-detector-networks.

\subsection{Network of 3 detectors}
\begin{figure}[htbp]
\begin{minipage}[c][15cm][t]{\textwidth}
%  \vspace*{\fill}
  \centering
  \includegraphics[width=\textwidth]{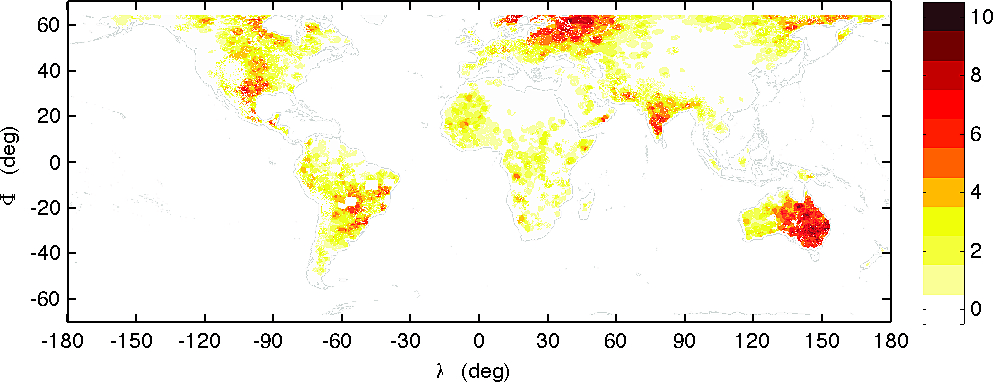}
  \subcaption{ }
  \label{fig:CD3_90}\par\vfill
  \includegraphics[width=\textwidth]{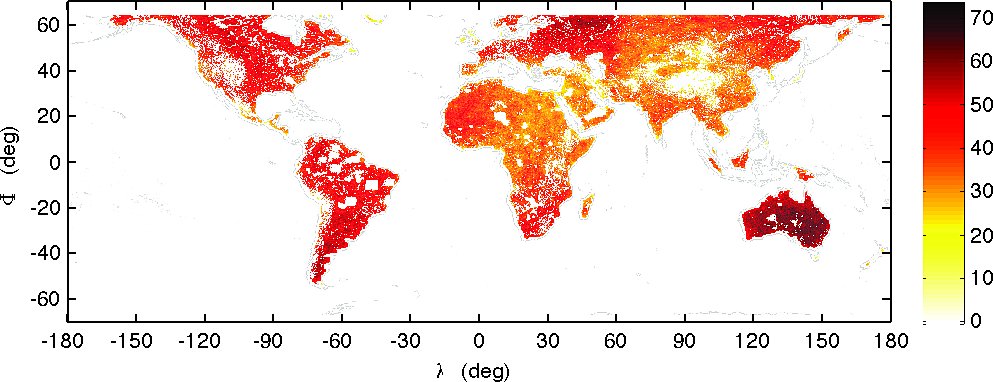}
  \subcaption{ }
  \label{fig:CD3_80}
\end{minipage}
\caption{World map showing the ``flexibility index" -- i.e. the number of different, distinct network configurations associated with a particular site, assuming a 3-detector GW detector network. The upper panel shows the result after filtering out all configurations with FoM smaller than $90\%$ of the most optimal configuration, while the lower panel shows the result with the filtering criteria set to be $80\%$.}\label{fig:CD3} 
\end{figure}

As shown in figure \ref{fig:CD3_90}, after filtering out all configurations that have a FoM smaller than $90\%$ of the highest FoM, the `best' site (as defined by the flexibility index introduced in the previous section) is located in Australia.
This result is consistent with previous conclusions from \cite{Raffai2013}.
For the best site identified in this way there are in total 10 different possible network configurations.
Notice that in some regions of Europe, North America and India there are also large numbers of alternative network configurations with high FoM, and Australia is only slightly better than these regions.

\begin{figure}
\begin{minipage}[c][9cm][t]{.5\textwidth}
  \vspace*{\fill}
  \centering
  \includegraphics[width=6cm]{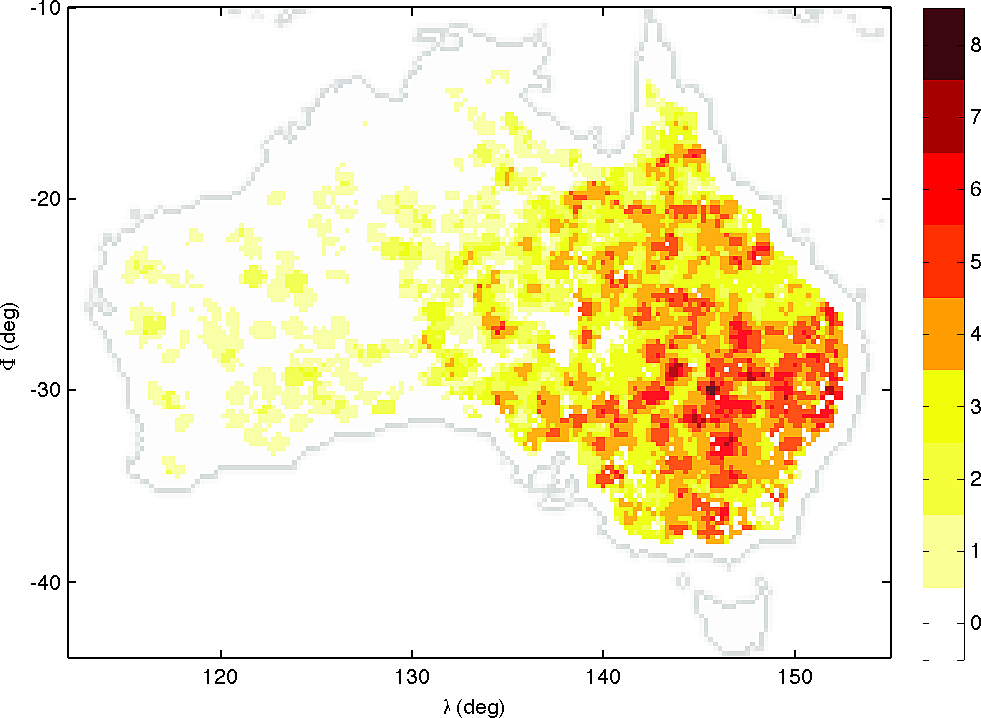}
  \subcaption{ }
  \label{fig:Aus3_90}
  \includegraphics[width=6cm]{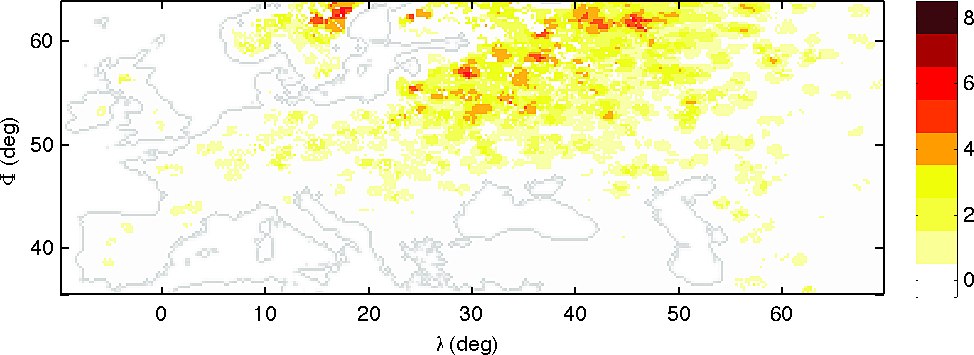}
  \subcaption{ }
  \label{fig:Eur3_90}
\end{minipage}%
\begin{minipage}[c][9cm][t]{.5\textwidth}
  \vspace*{\fill}
  \centering
  \includegraphics[width=6cm]{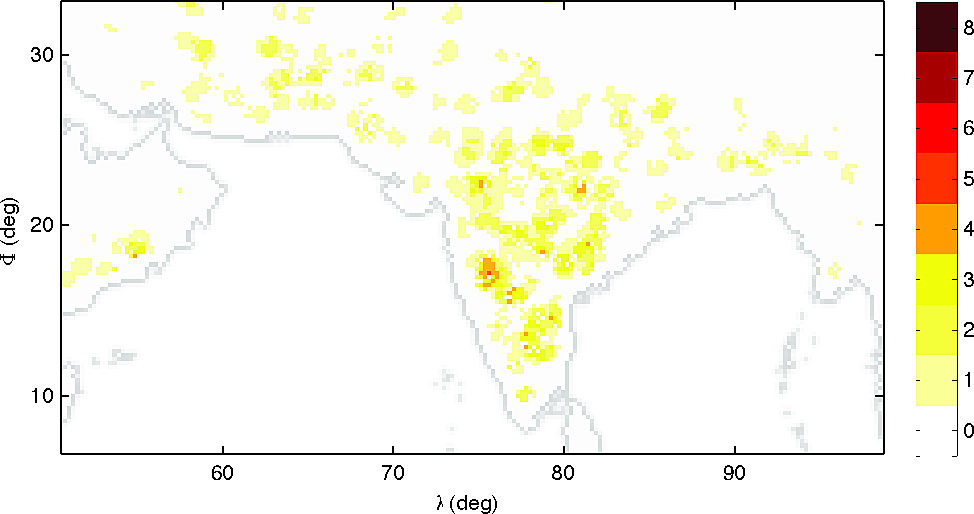}
  \subcaption{ }
  \label{fig:Ind3_90}\par\vfill
  \includegraphics[width=6cm]{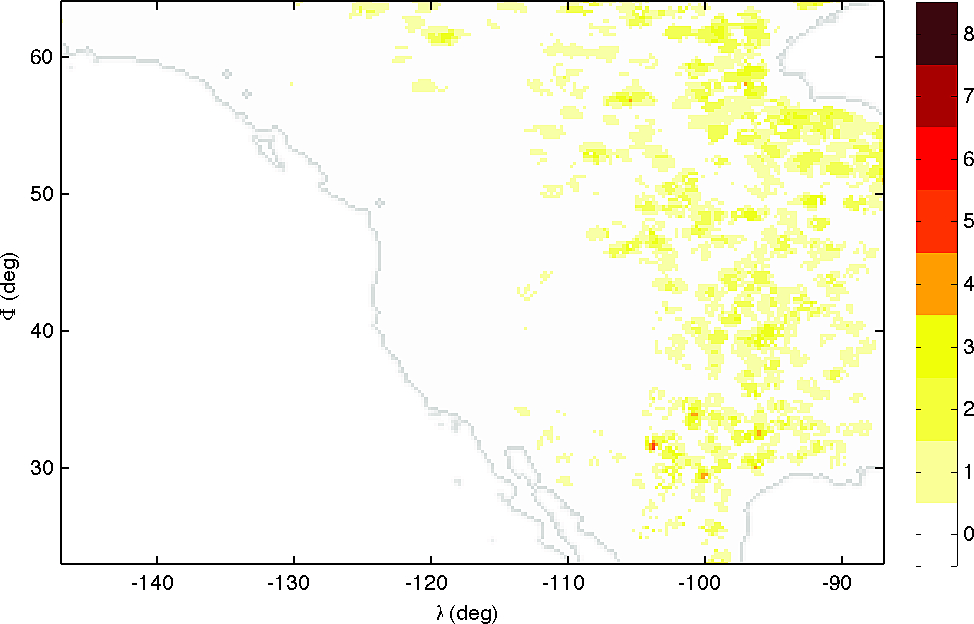}
  \subcaption{ }
  \label{fig:NAm3_90}
\end{minipage}
\caption{Zoom in map for regions with high flexibility indices, from figure \ref{fig:CD3_90}. Subplot a, b, c and d shows maps of Australia, Europe, India and America respectively. Notice that a shorter smooth length is applied here compared with figure \ref{fig:CD3}, causing a larger fluctuation.}
\label{fig:zoom3_90}
\end{figure}

\begin{figure}
\begin{minipage}[c][9cm][t]{.5\textwidth}
  \vspace*{\fill}
  \centering
  \includegraphics[width=6cm]{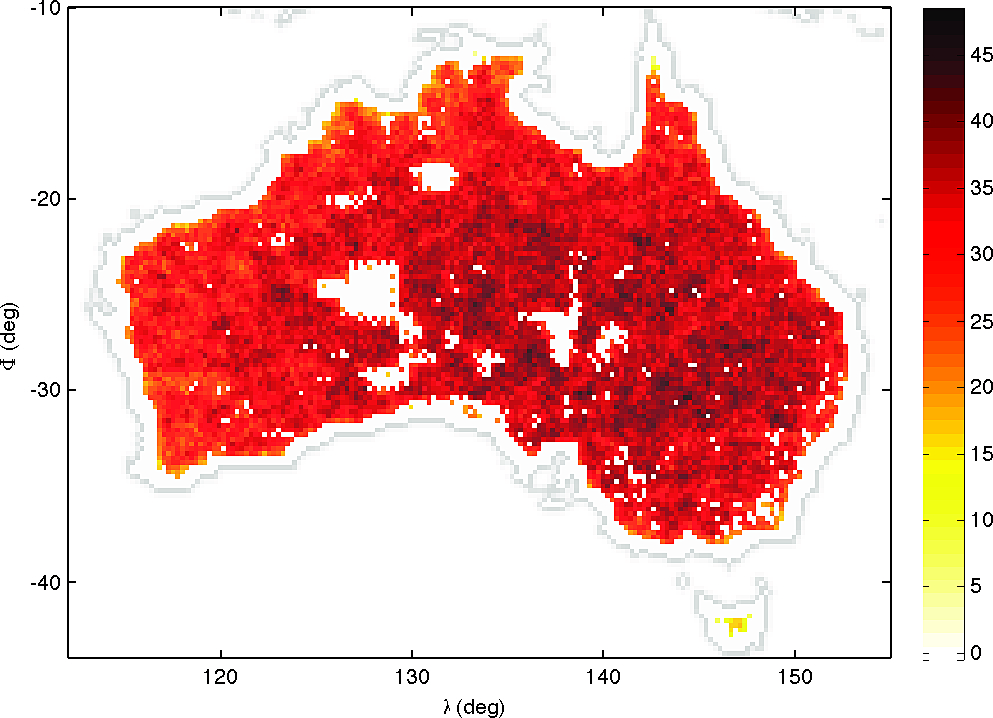}
  \subcaption{ }
  \label{fig:Aus3_80}
  \includegraphics[width=6cm]{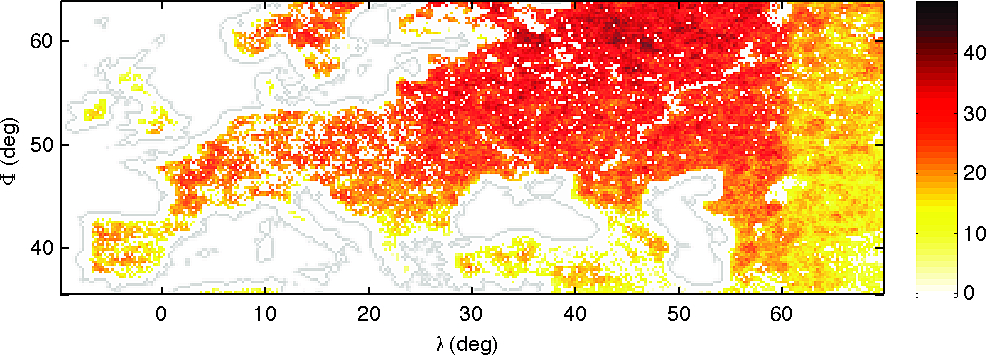}
  \subcaption{ }
  \label{fig:Eur3_80}
\end{minipage}%
\begin{minipage}[c][9cm][t]{.5\textwidth}
  \vspace*{\fill}
  \centering
  \includegraphics[width=6cm]{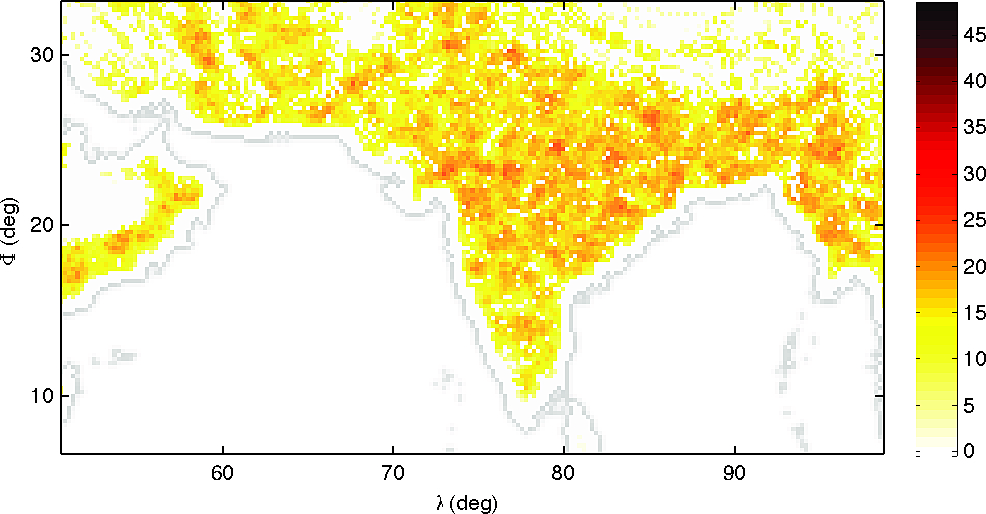}
  \subcaption{ }
  \label{fig:Ind3_80}\par\vfill
  \includegraphics[width=6cm]{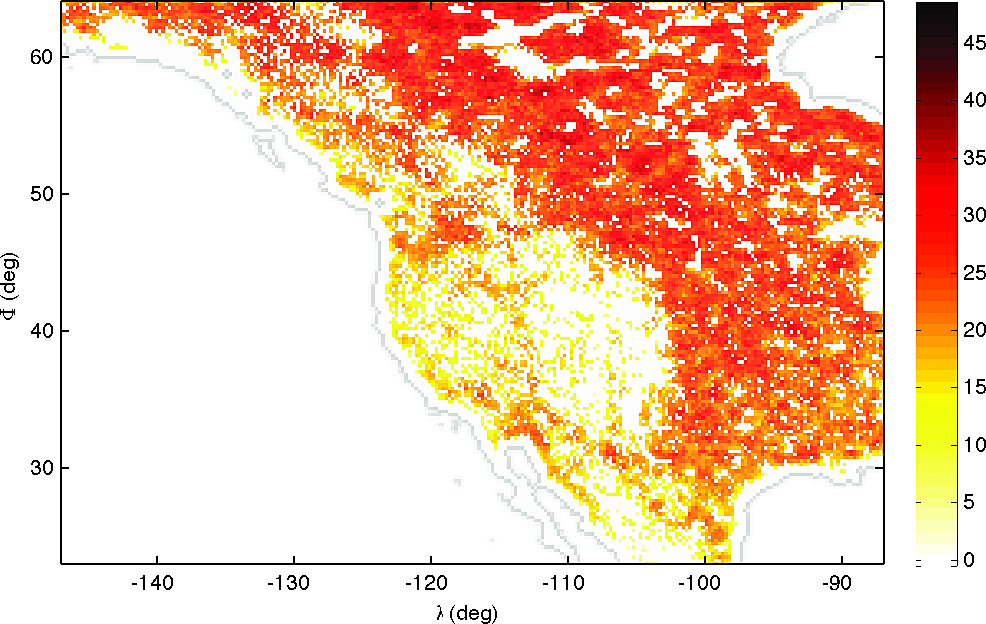}
  \subcaption{ }
  \label{fig:NAm3_80}
\end{minipage}
\caption{Zoom in map for regions in a 3-detector network with high flexibility indices, from figure \ref{fig:CD3_80}. Subplot a, b, c and d shows maps of Australia, Europe, India and America respectively. Notice that a shorter smoothing length is applied here compared with figure \ref{fig:CD3}, causing a larger fluctuation.}
\label{fig:zoom3_80}
\end{figure}

Figures \ref{fig:Aus3_90} to \ref{fig:NAm3_90} show zoomed-in detail on those regions around the world which have highest flexibility index: Australia, Europe, India and North America.

We also applied a brute force search on the `ideal Earth' (see \ref{sec:opt}) in order to check the optimal network configuration.
This exhaustive method can search for the optimal configuration of a 3-detector network when not including any terrestrial constraints. We find that such a network would be optimal when the detectors form an isosceles triangle with two sides of length of $\sim 130^\circ$, and the distinct third side of length $\sim 50^\circ$.
This result further confirms the previous result that when a network consists of only two detectors, the optimal situation is when they are separated by $\sim 130^\circ$ \cite{Raffai2013}.

We can also notice that Central Africa and East Asia seem not to be ideal sites.
The major reason for this is that the $130^\circ$ circle around these regions mostly falls in the ocean.

In figure \ref{fig:CD3_80}, the world map is shown for the case of a lower FoM threshold: here we have filtered out network configurations that have FoM less than $80\%$ of the optimal value.
Notice that some areas in Africa and in East Asia that were blank in the previous figure, with a $90\%$ threshold, are now filled in.
This indicates that the blank regions in figure \ref{fig:CD3_90} were not the result of insufficient sampling but rather were due to the
intrinsic lack of high FoM configurations in these regions.
It seems that these new potential interferometer sites are either too close to the sea, or are otherwise not ideal for building future generation detectors, however.

\begin{table}
\centering
  \input{table1.tex}
  \caption{
The locations of the other two detectors, for a series of examples of ``good" 3-detector networks, in which the future generation detector is located around the global optimal site, at longitude $146^\circ$ and latitude $-30^\circ$.  Each of these networks yields a value of $C$ that is greater than $90\%$ of $C_{max}$.
The order does not have any special meaning, so the site 1 and site 2 are interchangeable.
    \label{tab:3Dsites}}
\end{table}

In table \ref{tab:3Dsites} we list the locations of the other two sites for various ``good" 3-detector networks, in which one site is located near to the global optimal site.

\subsection{Network of 5 detectors}
\begin{figure}[htbp]
\begin{minipage}[c][15cm][t]{\textwidth}
%  \vspace*{\fill}
  \centering
  \includegraphics[width=\textwidth]{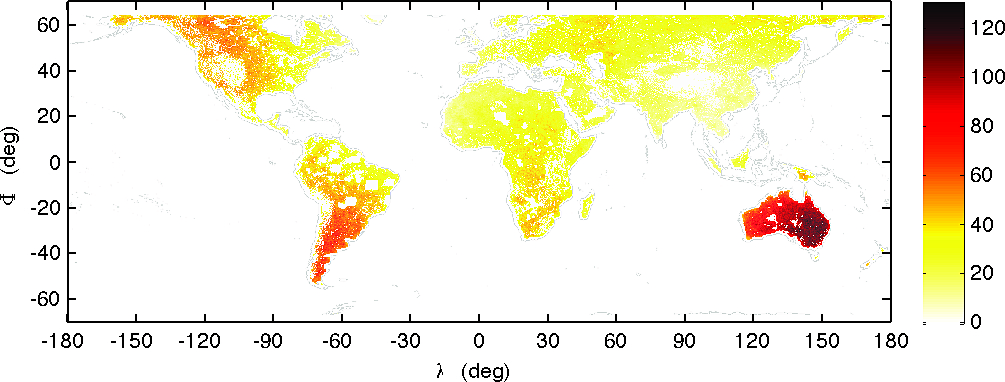}
  \subcaption{ }
  \label{fig:CD5_90}\par\vfill
  \includegraphics[width=\textwidth]{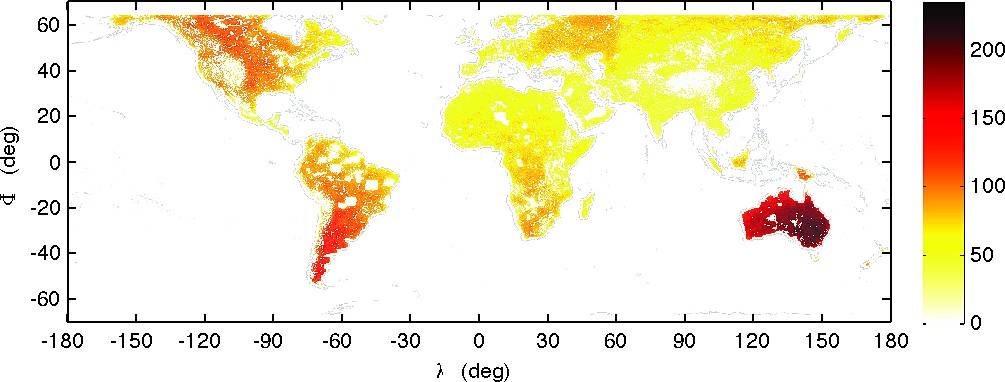}
  \subcaption{ }
  \label{fig:CD5_80}
\end{minipage}
\caption{World map showing the flexibility index, assuming a 5-detector GW detector network. The upper panel shows the result after filtering out all configurations with FoM smaller than $90\%$ of the optimal configuration, while the lower panel shows the result with the filtering criteria to be $80\%$.}\label{fig:CD5}
\end{figure}

Figure \ref{fig:CD5} shows a world map of the flexibility index for a 5-detector network.
Here we find that Australia is still the best site, and unlike the situation with a 3-detector network, it is significantly better than any other regions.
In the best site, there are in total 131 different possible network configurations after filtering out all configurations with FoM smaller than $90\%$ of the optimal FoM.
Lowering the threshold to $80\%$ increases this number to 235.
Besides Australia, the next best site locations are in North America and South America, although the flexibility index for these locations is more than $50\%$ smaller than for Australia.

Another important respect in which our 5-network results differ from those of the 3-detector network is that the first detector can be built almost anywhere, as long as it is not excluded by the conditions described in section~\ref{sec:exc}.
One should not be surprised by this outcome since the more detectors that a network includes the more flexible it should become.

\begin{figure}
\begin{minipage}[c][9cm][t]{.5\textwidth}
  \vspace*{\fill}
  \centering
  \includegraphics[width=6cm]{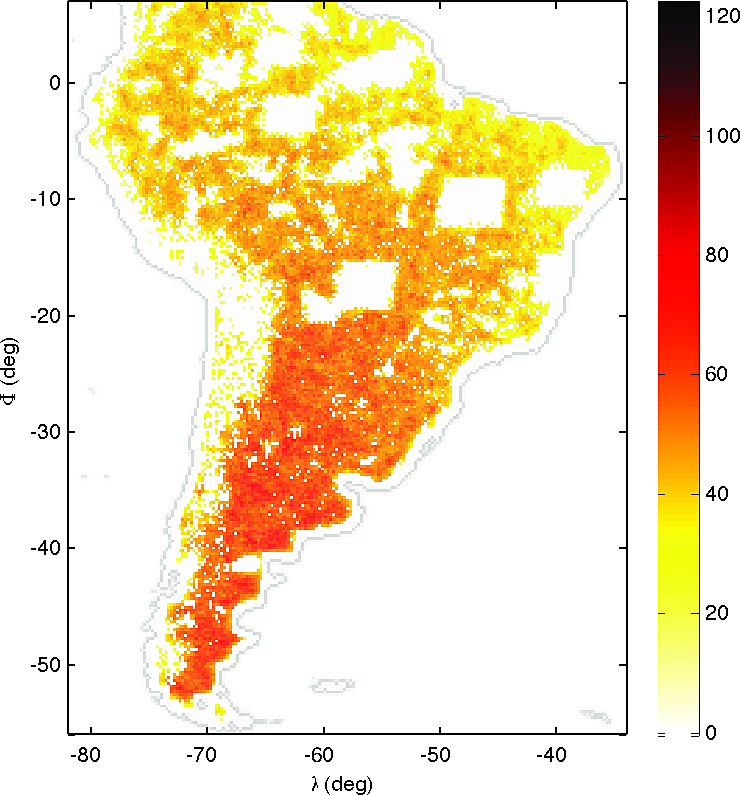}
  \subcaption{ }
  \label{fig:SAm5_80}
\end{minipage}%
\begin{minipage}[c][8cm][t]{.5\textwidth}
  \vspace*{\fill}
  \centering
  \includegraphics[width=5cm]{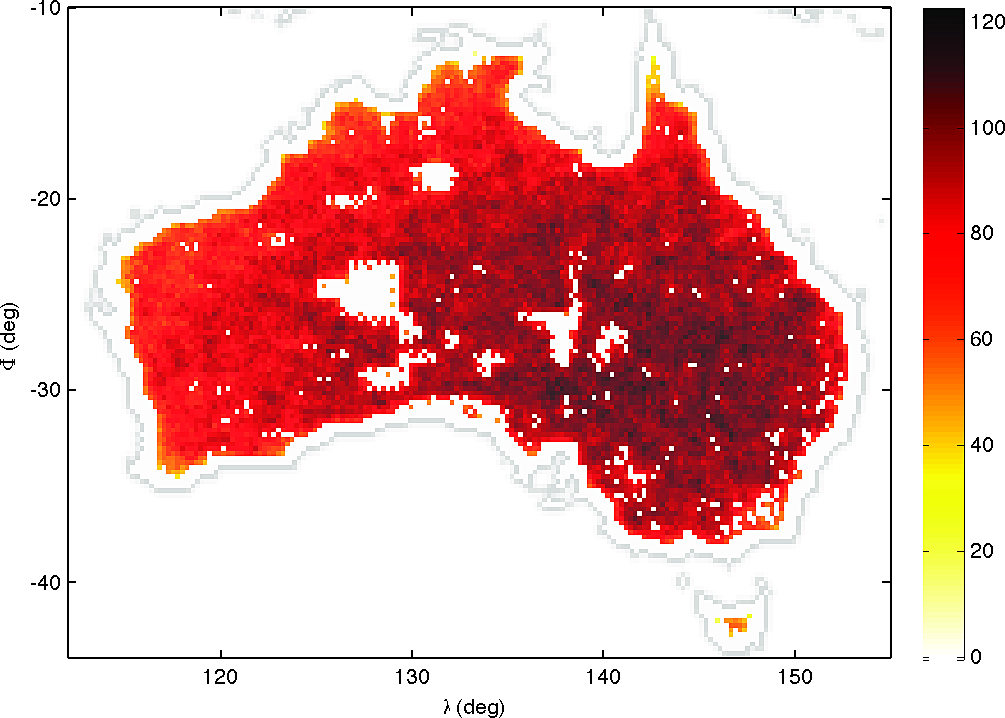}
  \subcaption{ }
  \label{fig:Aus5_80}\par\vfill
  \includegraphics[width=5cm]{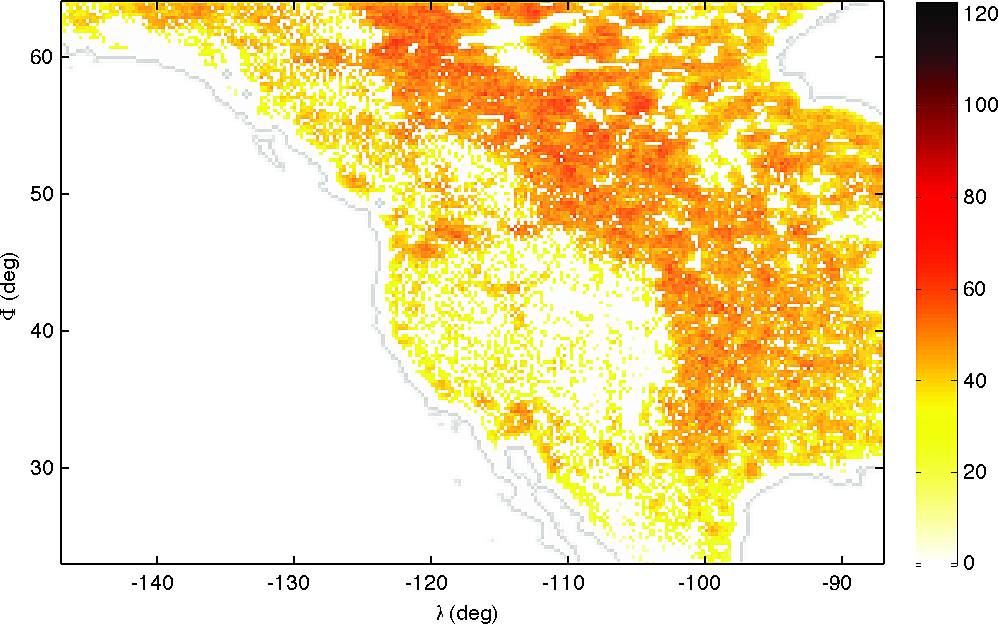}
  \subcaption{ }
  \label{fig:NAm5_80}
\end{minipage}
\caption{Zoom in map for regions in a 5-detector network with high flexibility indices, from figure \ref{fig:CD5_80}. Subplot a, b, and c shows maps of South America, Australia and America respectively. Notice that a shorter smoothing length is applied here compared with figure \ref{fig:CD5}, causing a larger fluctuation.}
\label{fig:zoom5_80}
\end{figure}

Figures \ref{fig:SAm5_80} to \ref{fig:NAm5_80} show zoomed-in detail on those regions around the world which have highest flexibility index for a 5-detector network, and adopting a FoM filtering threshold of $80\%$.

\begin{table}
\centering
  \input{table2.tex}
  \caption{
The locations of the other four detectors, for a series of examples of ``good" 5-detector networks, in which the $5^{th}$ detector is located around the global optimal site, at longitude $146^\circ$ and latitude $-29^\circ$. These examples are the networks that give the 10 largest sampled values of $C$.
The order does not have any special meaning, so the four sites in one line are interchangeable.
    \label{tab:5Dsites}}
\end{table}

In table \ref{tab:5Dsites} we list the locations of the other four sites for various ``good" 5-detector networks, in which one site is located near to the global optimal site.

\section{Conclusions and Discussion}

In this work we have applied a novel method for identifying optimal sites for future networks of Gravitational Wave detectors.
Our method adopted a new sampling approach that is well-suited to dealing with high-dimensional parameter spaces, thus permitting for the first time the simultaneous optimization of parameters for multi-detector networks -- a significant extension of the method previously presented in \cite{Raffai2013}.
We presented results for networks comprising 3 and 5 GW detectors.

We adopted a FoM for each network configuration based on its capability of reconstructing the signal polarization, accuracy of source localization and accuracy of source parameters estimation for a standard compact binary source. 
We followed the definition previously adopted in \cite{Raffai2013} for the combined FoM, $C$; however, the actual definition of $D$ in this work is slightly different, as we calculate the fraction of sky for which the source can be localized to better than a specified area, $S$.

We used a Bayesian, MCMC-based sampling method which meets our requirement to sample efficiently in high-dimensional parameter spaces.
However, for a multi-modal posterior distribution standard MCMC is much less effective.
We have, therefore, developed a variant, known as mixed MCMC, that is ideally suited to handle the multi-modal feature of our problem.
We partition the parameter space manually to facilitate fast initial sampling, dividing the world into 6 regions, roughly overlapping with the normal definition of the continents.
The multiple modes of our distribution are expected to be located distinctly in combinations of these 6 regions.
Such a partition is somewhat arbitrary, but this is intuitive, and the purpose is to distinguish all modes so that no two modes share one piece of the partition.
As long as the partition is dense enough, we should obtain conclusions that are robust against changes in the partition. 

The sampling results were combined and for each pixel in a $1520\times759$ pixel world map (corresponding to a resolution of approximately $26$ km at the equator), we counted the number of distinct network configurations that have a FoM higher than $90\%$ of the best FoM identified.
We call this number the ``flexibility index" of the network: in this sense, a site with a large flexibility index offers more options for network configurations with a high FoM.
In other words, once a detector is built on such a site, one has greater flexibility for locating other detectors.
This criterion to identify a ``good" detector location is, therefore, well suited to the (likely) case where future detector networks are not built simultaneously but sequentially.

For both 3-detector and 5-detector networks, we consistently found that Australia hosted the best site -- further confirming and generalizing the conclusions of previous work in \cite{Raffai2013}, where only one detector was optimized.
However, for the 3-detector network, the best sites in Australia are only slightly better (in terms of their flexibility index) than optimal sites in Europe, America and India.
For a 5-detector network, on the other hand, Australia is a considerably better site than any other region.
This would suggest that, if the long-term goal is to create a network of as many as 5 GW detectors, then building one of the first detectors in Australia is a powerful strategy.

We have included two tables (in table \ref{tab:3Dsites} and in table \ref{tab:5Dsites}) showing the example locations of sites -- in networks of 3 GW detectors and 5 GW detectors respectively -- which, when combined with a detector located at the global optimal site, yield a combined FoM $C$ that is more than 90\% of $C_{max}$.    

Our approach is simplified in several important respects -- not least our assumption that the spatial distribution of GW sources is isotropic, and the seismic environment is homogeneous within the allowable regions, so that the comparison of different networks is purely determined by their relative geometric shape and (provided that all sites remain within allowed regions on the Earth's surface) is insensitive to translation of the entire network configuration.
The definition of the FoM could be improved to include such factors as economic stability and scientific policies.
Given the difficulty in modelling them accurately, especially over a timescale of decades, we have not considered such factors in this work.
The final decision of such a site selection would have to account for them.
Nontheless, we worked out a solid framework that allow future updates and the inclusion of any emerging geopolitical, military, financial, etc. constraints.
We are planning to create a crowdsourced project to update and refine the constraint map with all available information, so that we can keep the boundary conditions up-to-date and detailed.

In the future, we can include phase information of GW waveform into considertaion; arrange the weights of individual FoMs to be unbiased as some FoM could be more sentive to configurations than others; correlation between FoMs should be investigated, since for future generation detectors, the ability of distinguish polarization could be used to help constraining sky localization \cite{Merkowitz1995} \cite{Merkowitz1997}.
In future work we will also extend our approach to include astronomical information about the actual, anisotropic, distribution of potential GW sources on the sky;  this information will break the degeneracy of our FoM to network translation.
Meanwhile, seismic stability for a site can also be considered quantitatively in the FoM of the network, further providing more realistic optimization.

\section*{Acknowledgements}
The authors want to thank Stanley for pushing the application of MCMC to this project, and we are grateful for the priceless comments from Stefan Countryman and Alexa Staley, and the discussion with Stanley, Sathyaprakash, Alan is really inspiring.
The authors would also like to thank the anonymous referees for valuable comments and suggestions.
Yiming Hu was supported by the China Scholarship Council and Principal's Early Career Mobility Fund.
Peter Raffai assisted in fundamental research within the framework of T\'AMOP 4.2.4. A/1-11-1-2012-0001 National Excellence Program 
``Elaborating and operating an inland student and researcher personal support system", was realized with personal support.
The project was subsidized by the European Union and co-financed by the European Social Fund.
The authors are grateful for the support from the Science and Technology Facilities Council of the United Kingdom, the Scottish Universities Physics Alliance and Columbia University in the City of New York.
This work has been supported by the United States National Science Foundation under cooperative agreement PHY-0847182

\appendix
\section{Constructing the partition}\label{sec:part}
\begin{figure}[htbp]
\begin{center}
\includegraphics[width=\textwidth]{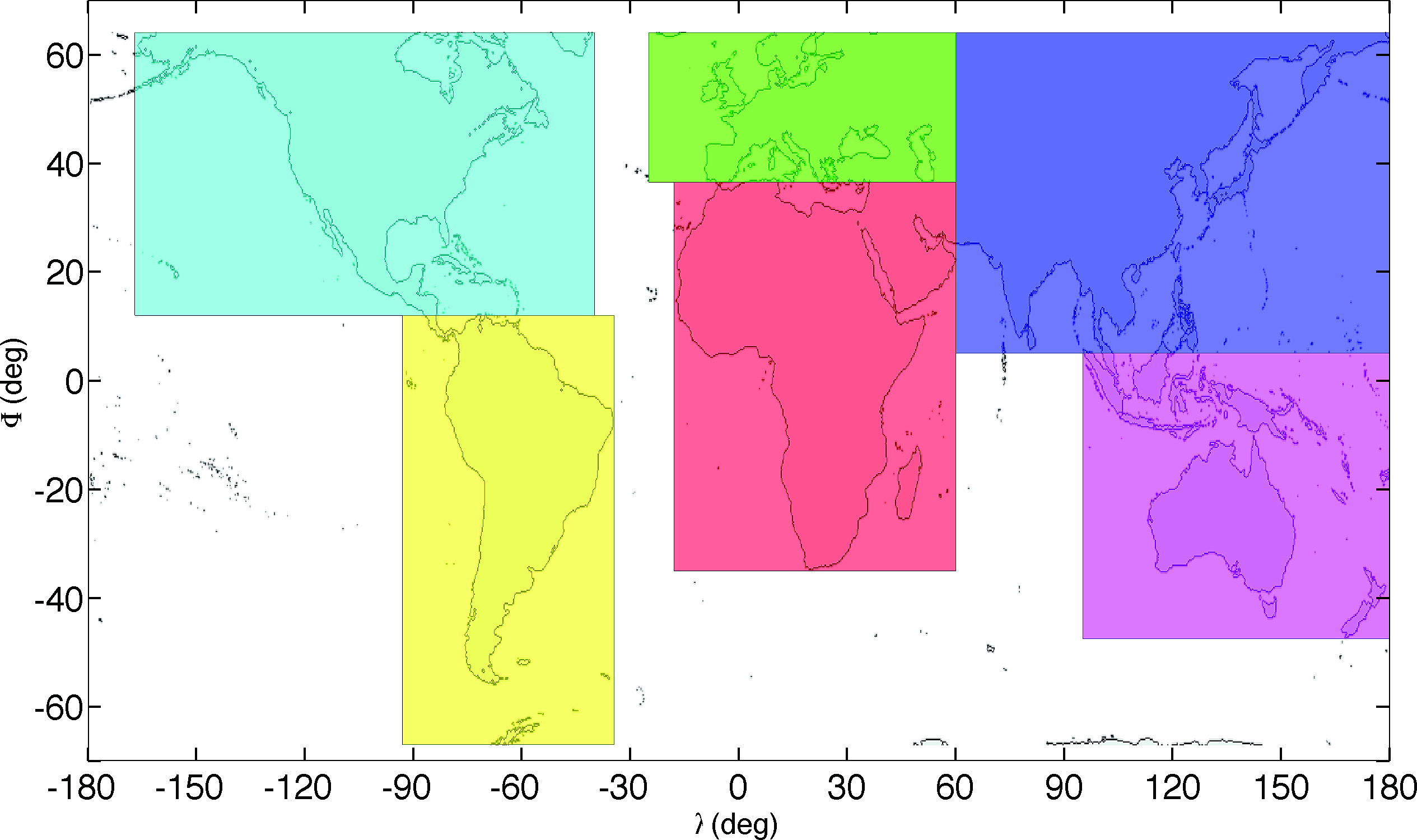}
\caption{Division of the world map into 6 distinct sections, roughly speaking coincident with the 6 populated continents: North America, South America, Europe, Africa, Asia and Australia. Notice that, although there are some areas that are not covered by any section, these areas are mostly islands and consequently all will be rejected either by the criterion of lying in a polar region or the criterion of being less than 100 km inland.}
\label{fig:div}
\end{center}
\end{figure}

In our original application of mixed MCMC in {\em Hu et al.\/}  \cite{Hu2014}, the multiple modes were assumed to have been identified using methods such as parallel tempering, so that the identification is achieved objectively.
However, in the application considered here we can further simplify this process by manually partitioning the allowable regions into 6 patches corresponding to the Earth's continents -- i.e. North America, South America, Europe, Africa, Asia and Australia, as shown in figure \ref{fig:div}.

We assume that for $n$ detectors, each of them can be located in one of these 6 continents. 
If one continent can only host up to one detector, then the number of different possible combinations is just simply $C^n_6$.
In the case of a 3-detector network, this number is $C^3_6=20$.
However, we should also consider that some continents like Asia are very large in area, so that locations in one continent can be a considerable distance apart.
In order to be conservative, therefore, we enable each continent to host more than one detector.
Hence we must include among the possible configurations those networks for which one continent hosts two detectors; in total this makes $C^1_6C^1_5=30$ cases, together with the $C^1_6=6$ cases in which all detectors are located in one continent.
So for the 3-detector network, there are in total 56 different possible combinations of host continents; in each such combination, one sub-chain of mixed MCMC is assigned to sample and we identify combinations with sub-chains.
Analysis of the 5-detector network situation is identical in principle, although somewhat more complicated: it is straightforward to show that there are in total 252 different combinations in this case.

Notice the assumption that is implied here, that there will be generally no more than 1 local posterior peak in one subchain.
We believe that this assumption is sound and justified, and it can help us vastly to simplify the process of initializing the chains.

\section{Optimizing parameters}\label{sec:opt}
Originally, the calculation of FoM for one single network configuration would take $\sim100$s -- clearly emphasizing the need for optimization of a conventional MCMC requiring of order $10^5$ samples.

First, the original calculation of the $I$ and $R$ FoM was achieved through numerically integrating over the whole sky with a very high angular resolution.
We then tested empirically the relation between sky resolution and calculation accuracy using a detector network on an hypothetical idealized Earth with the first detector site fixed to be at the North Pole, the second set fixed along the Prime Meridian of longitude and the future site placed uniformly over the surface of the sphere.
Because of the symmetry inherent in our FoM, any actual multi-detector network can be rendered equivalent to this idealized network through appropriate choice of coordinate system.

\begin{figure}[htbp]
\begin{center}
\includegraphics[width=\textwidth]{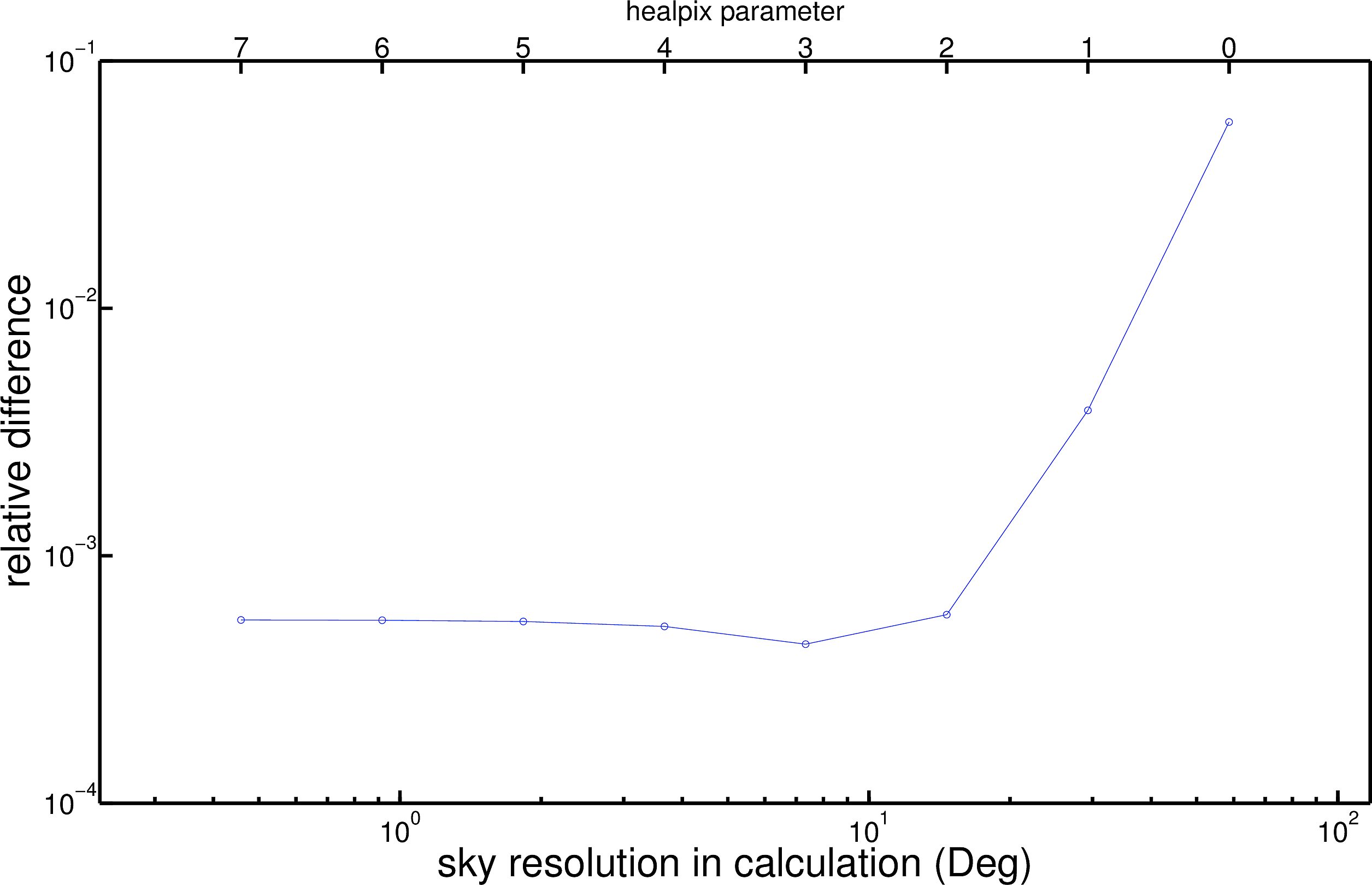}
\caption{Relative change of the $I$ FoM, averaged over all possible combinations from the `hypothetical Earth' as described in the text, as a function of \emph{healpix} input parameter.
As expected, when the resolution is made smaller, the relative difference becomes smaller, while the uncertainty also becomes smaller, which explains the minimum value in the \emph{healpix} parameter 3.}
\label{fig:rvc}
\end{center}
\end{figure}

To calculate $I$ or $R$, we first discretize the whole sky into uniformly distributed representative points using the \emph{healpix} algorithm \cite{Gorski2005}, and calculate corresponding individual FoMs for a source located at each of these points.
The input to \emph{healpix} is a positive integer that determines the resolution: the higher this input parameter, the more points we discretize and the more accurate $I$ and $R$ will be -- although the calculations will also be more time-consuming.
Since the actual number of calculations is proportional to the square of the resolution, the gain in computational efficiency from reducing this \emph{healpix} parameter is huge.
We calculated the individual $I$ and $R$ FoMs for a range of different resolutions, and compared them with the values obtained using the highest resolution that was feasible -- which we defined as that obtained for a \emph{healpix} parameter of 8.
The relative difference in the FoMs, averaged over all combinations of networks from the aforementioned hypothetical Earth, was then calculated for smaller values of the \emph{healpix} input parameter.
As shown in figure~\ref{fig:rvc}, a much lower resolution (with a \emph{healpix} parameter of 3) will only result in a negligible loss of accuracy in the average FoMs.
When the resolution is increased, variance is expected to increase, causing the non-monotonic behaviour seen in figure~\ref{fig:rvc}.
We therefore adopted a \emph{healpix} parameter of 3 in our subsequent analysis, corresponding to an angular resolution of 0.13 radian, or 7 degrees.
We understand this choice as a natural outcome of the fact that 7 degrees is much smaller than the characteristic angular length for the antenna pattern to vary.

\begin{figure}[htbp]
\begin{center}
\includegraphics[width=\textwidth]{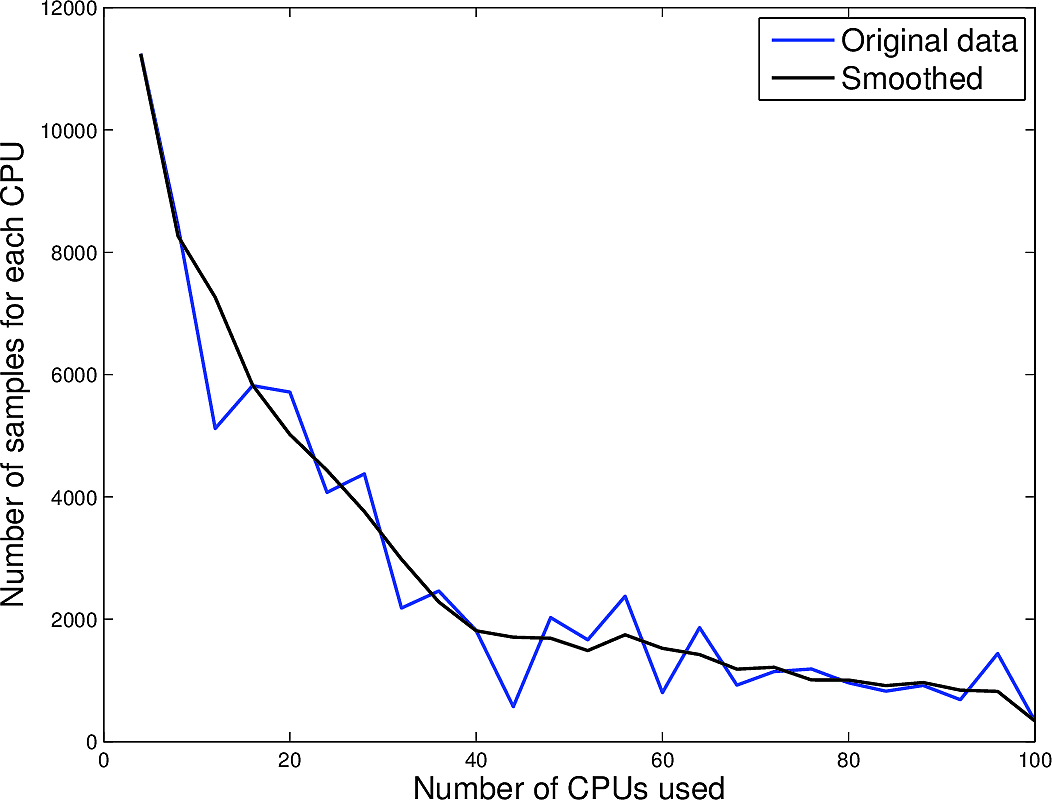}
\caption{Average number of samples for a 3-detector network using different numbers of CPUs. Since the Gelman-Rubin criterion was applied on 4 major chains, the tested numbers are multiples of 4 (for details see section \ref{sec:realis}). We can see that there is a uniform decrease in the number of samples as more CPUs are used. We apply a moving average filter with 5 points to obtain the smoothed data. The vertical axis is just for illustration, and does not correspond to the actual number of samples obtained, since we apply different stopping criteria in this test compared with that used in our actual sampling. However our conclusion about the optimal number of CPUs suitable for our analysis should remain valid.}
\label{fig:avg}
\end{center}
\end{figure}

Secondly we explored the optimal number of CPUs to use.
It is natural to expect that using more CPUs should lead to greater computational efficiency, and this was one of the main motivations for the development of mixed MCMC.
However, we can expect that there will be a limiting behaviour such that, for a sufficiently large number of CPUs, further increasing this number will not result in any significant further improvement, since every sampler needs some certain sample time to get some reliable estimation of the total parameter space.
We illustrate this trend in figure~\ref{fig:avg}, where we see that the improvement in efficiency reduces substantially after the number of CPUs exceeds $\sim 40$.
Thus we determined the optimal number of CPUs to be 40 when sampling for a 3-detector network.
We assumed further that this optimal number should be proportional to the total number of distinct network configurations, as described in \ref{sec:part}.
Hence for a 5-detector network we adopted as the optimal choice of CPU number $40\times252/56\sim 200$.
Our numerical test runs showed that for a 5-detector network, using 200 CPUs did indeed sample much faster than using 40 CPUs, while still yielding satisfactory results.

Also, we investigated the usage of Gelman-Rubin criterion \cite{Gelman1992}.
When multiple realizations of the same model are running simultaneously, the properties of MCMC guarantee that different realizations should give a similar distribution after a sufficiently long time.
The variance of each parameter within each MCMC realization was calculated, and the estimation of the intrinsic variance was constructed with the information of multiple realizations. 
The ratio between this estimate of the intrinsic variance and the variance within each realization is thus calculated; this is defined as the Gelman-Rubin $R$ value.
Due to the way it is constructed, the $R$ value on every parameter always tends to be larger than unity, but when the MCMC chain has converged it will asymptotically tend towards $R=1$.

In this problem, we set a Gelman-Rubin criterion of $R=1.1$ -- that is, if no parameter in a given subchain yields a Gelman-Rubin $R$ value larger than 1.1, we will label this subchain as ``converged''.
However, the sampling in this sub-chain is continued, otherwise the property of detailed balance will be totally destroyed and we can't predict its impact on our sampling result.
In order to compute the Gelman-Rubin $R$ for $m$ different MCMC realizations of $n$ points $x_i$, one needs to construct
\begin{equation}\label{eq:B}
B = n\sum_{i=1}^m (\bar{x_i}-\bar{x})^2/(m-1)
\end{equation}
where $\bar{x_i}$ and $\bar{x}$ are the mean of each MCMC and total sample, separately.
\begin{equation}\label{eq:W}
W = \sum_{i=1}^{m} s_i^2/m
\end{equation}
is the average of the variances defined as $s_i^2$.
Based on these values, one can compute
\begin{equation}\label{eq:sigma}
\hat{\sigma}^2 = \frac{n-1}{n}W + \frac{B}{n}
\end{equation}
\begin{equation}\label{eq:V}
\sqrt{\hat{V}} = \sqrt{\hat{\sigma}^2+B/mn}
\end{equation}
\begin{equation}\label{eq:df}
dof = 2\hat{V}^2/\hat{var}(\hat{V})
\end{equation}
for the estimation of the target variance, the estimation of the sampling variance and the number of degrees of freedom respectively, with
\begin{dmath}
\hat{var}(\hat{V}) = \Big(\frac{n-1}{n}\Big)^2\frac{1}{m}\hat{var}(s_i^2)+\Big(\frac{m+1}{mn}\Big)^2\frac{2}{m-1}B^2\\
\qquad+2\frac{(m+1)(n-1)}{mn^2}\frac{n}{m}[\hat{cov}(s_i^2,\bar{x_i^2})-2\bar{x}\hat{cov}(s_i^2,\bar{x_i})]
\end{dmath}
and the $R$ value is defined as $\sqrt{R} = \sqrt{(\hat{V}/W)dof/(dof-2)}$.
As $dof$ tends to infinity, the term $dof/(dof-2)$ will cancel out for large $n$.
So $R-1 \sim \frac{m+1}{m-1}\frac{\sum_{i=1}^m (\bar{x_i}-\bar{x})^2}{\sum_{i=1}^{m} s_i^2}$.

Note that in Bayesian parameter estimation, a conventional choice for the threshold to be converged is $R-1\le0.01$.
However, in our case, in order to accelerate the sampling process we run in parallel $4\times N$ chains and then sum the $N$ chains up to make up 4 ``major chains''.
The Gelman-Rubin $R$ criterion is then applied to these 4 major chains and a relatively large criterion $R-1\le0.1$ is adopted.
So for 3-detector-network, the stop criterion translates as meaning that the variance of average should not be larger than $0.08$ times of the average variance within groups, and for 5-detector-networks $0.09$ times.

The sampling process is then continued until all subchains have converged. 
We expect that the results would not be changed significantly if we were to adopt a stronger constraint on the $R$ criterion.
However, smaller $R$ criterion value would largely increase the computational cost.
The choice of $R=1.1$ is an sufficiently tight constatint while resulting in an acceptable computational burden.

Finally consider the choice of value of $S$, which is required to calculate $D$.
Notice that according to \cite{Fairhurst2010}, sky localization error is strongly related to SNR, which for a detection criterion is chosen to be 8 in general cases.
In previous studies, a typical event's localization error will span an area of $10-100$deg$^2$ with 2-3 GW detectors, and $\sim 5$deg$^2$ for a 5 detector network (e.g \cite{Fairhurst2010}).
Besides, the rule of thumb for chooing proper $S$ is to make the FoM is as discriminatory as possible.
If we set $S$ too small, then most networks will give a value barely larger than zero, and the influence of calculation uncertainty will be severe.
If, however, $S$ is too large then in almost all networks the fraction of the sky for which a source can be localized to a region smaller than $S$ also becomes very large, and once again we will lose the ability of the FoM to distinguish effectively between different network configurations.

For the 3-detector network, we find that $S$ is best set to $\approx$ 59.5 deg$^2$, in this case, the best configuration gives a $D$ close to but not equal to 1, thus avoiding the aforementioned degeneracy.
For the 5-detector network, however, $S$ is set to the much smaller value of $S=2.5$ area units, translating to $4.5$ 
deg$^2$. This is not surprising since it is generally expected that networks with more detectors will perform better with regard to this FoM.
Tests on an ideal Earth, as described in the previous sub-section, show that adopting the value of $S=2$ will lead to a significant degeneracy at $D=0$, while a value of $S=3$ leads to severe degeneracy at $D=1$.
Hence the choice of $S=2.5$ area units represents an appropriate trade-off between these two cases, which is equivalent to $4.5$ deg$^2$.
These values shows good consistency with \cite{Fairhurst2010}.

Notice that the ability of sky localization is closely related to the EM follow-up.
Since the sky localization area tends to be much bigger than ordinary telescopes' fields of view, one would be much more interested in the value of sky localization area, but not so interested in the shape or topology of the localization.
Thus we have good reason to believe that our definition of $D$ is reasonable and representitive.

\section*{References}
\bibliographystyle{unsrt}
\bibliography{FutureNet_jpg}

\end{document}

%% file: table1.tex
\begin{tabular}{c c|c c}
\multicolumn{2}{c|}{Site 1}&\multicolumn{2}{c}{Site 2}\\\hline
long($^\circ$)&lat($^\circ$)&long($^\circ$)&lat($^\circ$)\\\hline
 -84 &   14 &  -43 &  -15\\
 27  &  51  &  -99 &   62\\
-105 &   57 &  -76 &    1\\
 25  &   1  & -103 &   37\\
 27  &  20  &  17  &  64 \\
 53  &  62  &  -69 &  -16\\
 117 &    3 &  -92 &   47\\
 16  &  -3  &  -43 &  -20\\
 15  &  -8  &  21  &  45 \\
\end{tabular}

%% file: table2.tex
\begin{tabular}{c c|c c|c c|c c}
\multicolumn{2}{c|}{Site 1}&\multicolumn{2}{c|}{Site 2}&\multicolumn{2}{c|}{Site 3}&\multicolumn{2}{c}{Site 4}\\\hline
long($^\circ$)&lat($^\circ$)&long($^\circ$)&lat($^\circ$)&long($^\circ$)&lat($^\circ$)&long($^\circ$)&lat($^\circ$)\\\hline
18 &   -4 &   32 &   59 &  -109&  30 &  -71 &  -49\\
23 &  -29 &  109 &   61 &   74 &   25 &  -93 &   44\\
19 &  -2  & 114  &  61  & -131 & 58  & -64  &  -5 \\
21 & -29  &  42  &  62  & 107  &  61  & -73  &   5 \\
24 &   -9 &  117 &   41 & -118&  37 &  -70 &  -42\\
32 &   -7 &   74 &   49 &  -88&  38 &  -69 &  -15\\
35 &  -11 &   46 &   24 & -119&  41 &  -63 &  -36\\
 4 &   32 &   19 &  -23 & -114&  40 &  -64 &  -36\\
20 &   -8 &  -122 &   44 &  -71& -45 &  -57 &  -13\\
23 &  -31 &   19 &   51 &  150 &   61 &  -95 &   30\\
\end{tabular}

%% file: FutureNet_jpg.bbl
\begin{thebibliography}{10}

\bibitem{Abbott2009}
B~P Abbott, R~Abbott, R~Adhikari, et~al.
\newblock Ligo: the laser interferometer gravitational-wave observatory.
\newblock {\em Reports on Progress in Physics}, 72(7):076901, 2009.

\bibitem{Accadia2012}
T~Accadia, F~Acernese, M~Alshourbagy, et~al.
\newblock Virgo: a laser interferometer to detect gravitational waves.
\newblock {\em Journal of Instrumentation}, 7(03):P03012, 2012.

\bibitem{Harry2010}
Gregory~M Harry and the LIGO Scientific~Collaboration.
\newblock Advanced ligo: the next generation of gravitational wave detectors.
\newblock {\em Classical and Quantum Gravity}, 27(8):084006, 2010.

\bibitem{Virgo2009}
The~Virgo Collaboration.
\newblock Advanced virgo baseline design, 2009.

\bibitem{LSCrate2010}
J~Abadie, B~P Abbott, R~Abbott, et~al.
\newblock Predictions for the rates of compact binary coalescences observable
  by ground-based gravitational-wave detectors.
\newblock {\em Classical and Quantum Gravity}, 27(17):173001, 2010.

\bibitem{ET2011}
the ET~science team.
\newblock Einstein gravitational wave telescope conceptual design study, 2011.

\bibitem{Sathyaprakash2012}
B~Sathyaprakash, M~Abernathy, F~Acernese, et~al.
\newblock Scientific objectives of einstein telescope.
\newblock {\em Classical and Quantum Gravity}, 29(12):124013, 2012.

\bibitem{Chassande-Mottin2010}
Eric Chassande-Mottin, Martin Hendry, Patrick~J. Sutton, and Szabolcs
  M\'{a}rka.
\newblock {Multimessenger astronomy with the Einstein Telescope}.
\newblock {\em General Relativity and Gravitation}, 43(2):26, April 2010.

\bibitem{Sathyaprakash2009}
B.S. Sathyaprakash and Bernard~F. Schutz.
\newblock Physics, astrophysics and cosmology with gravitational waves.
\newblock {\em Living Reviews in Relativity}, 12(2), 2009.

\bibitem{Coughlin2012}
M~Coughlin and J~Harms.
\newblock Seismic topographic scattering in the context of gw detector site
  selection.
\newblock {\em Classical and Quantum Gravity}, 29(7):075004, 2012.

\bibitem{Raffai2013}
Péter Raffai, László Gondán, Ik~Siong Heng, et~al.
\newblock Optimal networks of future gravitational-wave telescopes.
\newblock {\em Classical and Quantum Gravity}, 30(15):155004, 2013.

\bibitem{Schutz2011}
Bernard~F Schutz.
\newblock Networks of gravitational wave detectors and three figures of merit.
\newblock {\em Classical and Quantum Gravity}, 28(12):125023, 2011.

\bibitem{Fan2014}
XiLong Fan, Christopher Messenger, and Ik~Siong Heng.
\newblock {A Bayesian approach to multi-messenger astronomy: Identification of
  gravitational-wave host galaxies}.
\newblock {\em pre-print}, page~22, June 2014.

\bibitem{Cutler1994}
C~Cutler and \'{E}e Flanagan.
\newblock {Gravitational waves from merging compact binaries: How accurately
  can one extract the binary's parameters from the inspiral waveform?}
\newblock {\em Physical review D: Particles and fields}, 49(6):2658--2697,
  March 1994.

\bibitem{Veitch2012}
J.~Veitch, I.~Mandel, B.~Aylott, et~al.
\newblock {Estimating parameters of coalescing compact binaries with proposed
  advanced detector networks}.
\newblock {\em Physical Review D}, 85(10):104045, May 2012.

\bibitem{Somiya2012}
Kentaro Somiya.
\newblock Detector configuration of kagra–the japanese cryogenic
  gravitational-wave detector.
\newblock {\em Classical and Quantum Gravity}, 29(12):124007, 2012.

\bibitem{Fairhurst2014}
S~Fairhurst.
\newblock Improved source localization with ligo-india.
\newblock {\em Journal of Physics: Conference Series}, 484(1):012007, 2014.

\bibitem{Fairhurst2010}
Stephen Fairhurst.
\newblock Source localization with an advanced gravitational wave detector
  network.
\newblock {\em Classical and Quantum Gravity}, 28(10):105021, 2011.

\bibitem{Metropolis1953}
Nicholas Metropolis, Arianna~W. Rosenbluth, Marshall~N. Rosenbluth, Augusta~H.
  Teller, and Edward Teller.
\newblock Equation of state calculations by fast computing machines.
\newblock {\em The Journal of Chemical Physics}, 21(6):1087--1092, 1953.

\bibitem{Hasting1969}
W.~K. Hastings.
\newblock {Monte Carlo sampling methods using Markov chains and their
  applications}.
\newblock {\em Biometrika}, 57(1):97--109, April 1970.

\bibitem{Gregory2005}
Phil~C. Gregory.
\newblock {\em Bayesian Logical Data Analysis for the Physical Sciences: A
  Comparative Approach with Mathematica Support}.
\newblock Cambridge, Cambridge, 2005 (ISBN: 0-521-84150-X).

\bibitem{Skilling2004}
John Skilling.
\newblock Nested sampling.
\newblock {\em AIP Conference Proceedings}, 735(1):395--405, 2004.

\bibitem{Skilling2006}
J.~Skilling.
\newblock {Nested sampling for general Bayesian computation}.
\newblock {\em Bayesian Analysis}, 1(4):833--860, 2006.

\bibitem{Sivia2006}
D.~S. Sivia.
\newblock {\em Data Analysis: A Bayesian Tutorial}.
\newblock Clarendon (Oxford Univ. Press), Oxford, 2006 (ISBN: 0-19-851762-9 or
  0-19-851889-7 in paperback).

\bibitem{Punturo2010}
M~Punturo, M~Abernathy, F~Acernese, et~al.
\newblock The third generation of gravitational wave observatories and their
  science reach.
\newblock {\em Classical and Quantum Gravity}, 27(8):084007, 2010.

\bibitem{vanderSluys2008}
Marc van~der Sluys, Vivien Raymond, Ilya Mandel, et~al.
\newblock Parameter estimation of spinning binary inspirals using markov chain
  monte carlo.
\newblock {\em Classical and Quantum Gravity}, 25(18):184011, 2008.

\bibitem{Raymond2009}
V~Raymond, M~V van~der Sluys, I~Mandel, et~al.
\newblock Degeneracies in sky localization determination from a spinning
  coalescing binary through gravitational wave observations: a markov-chain
  monte carlo analysis for two detectors.
\newblock {\em Classical and Quantum Gravity}, 26(11):114007, 2009.

\bibitem{Farr2013}
Benjamin Farr, Vicky Kalogera, and Erik Luijten.
\newblock {Efficient Estimation of Highly Structured Posteriors of
  Gravitational-Wave Signals with Markov-Chain Monte Carlo}.
\newblock {\em arXiv preprint arXiv:1309.7709}, pages 1--14, September 2013.

\bibitem{Swendsen1986}
Robert~H. Swendsen and Jian~S. Wang.
\newblock {Replica Monte Carlo Simulation of Spin-Glasses}.
\newblock {\em Physical Review Letters}, 57(21):2607--2609, November 1986.

\bibitem{Feroz2009}
F.~Feroz, M.~P. Hobson, and M.~Bridges.
\newblock Multinest: an efficient and robust bayesian inference tool for
  cosmology and particle physics.
\newblock {\em Monthly Notices of the Royal Astronomical Society},
  398(4):1601--1614, 2009.

\bibitem{Feroz2013}
F~Feroz, M~P Hobson, E~Cameron, and A~N Pettitt.
\newblock {Importance Nested Sampling and the MultiNest Algorithm}.
\newblock {\em arXiv preprint arXiv:1306.2144}, June 2013.

\bibitem{Hu2014}
Yi-Ming Hu, Martin Hendry, and Ik~Siong Heng.
\newblock {Efficient Exploration of Multi-Modal Posterior Distributions}.
\newblock {\em arXiv preprint arXiv:1408.3969}, August 2014.

\bibitem{Gelman1992}
Andrew Gelman and Donald~B. Rubin.
\newblock {Inference from Iterative Simulation Using Multiple Sequences}.
\newblock {\em Statistical Science}, 7(4):457--472, 1992.

\bibitem{agwed}
Aster gdem worldwide elevation data map.
\newblock \url{http://asterweb.jpl.nasa.gov/gdem.asp}.

\bibitem{natear}
Natural earth.
\newblock \url{http://www.naturalearthdata.com}.

\bibitem{night}
Nasa visible earth.
\newblock \url{http://visibleearth.nasa.gov/view.php?id=55167}.

\bibitem{prot}
World database on protected areas.
\newblock \url{http://protectedplanet.net/}.

\bibitem{tectonic}
{\em Exclusion based on tectonic plate lines.}

\bibitem{Merkowitz1995}
Stephen Merkowitz and Warren Johnson.
\newblock {Spherical gravitational wave antennas and the truncated icosahedral
  arrangement.}
\newblock {\em Physical review D: Particles and fields}, 51(6):2546--2558,
  March 1995.

\bibitem{Merkowitz1997}
Stephen Merkowitz and Warren Johnson.
\newblock {Techniques for detecting gravitational waves with a spherical
  antenna}.
\newblock {\em Physical Review D}, 56(12):7513--7528, December 1997.

\bibitem{Gorski2005}
K.~M. Górski, E.~Hivon, A.~J. Banday, et~al.
\newblock Healpix: A framework for high-resolution discretization and fast
  analysis of data distributed on the sphere.
\newblock {\em The Astrophysical Journal}, 622(2):759, 2005.

\end{thebibliography}
